\PassOptionsToPackage{dvipsnames}{xcolor}
\documentclass[english,twocolumn,a4paper]{article}

\usepackage[utf8]{inputenc}
\usepackage[big]{dgruyter}
\usepackage{microtype}

\usepackage{amsmath}
\usepackage{amssymb}
\usepackage{mathtools}
\usepackage{bm}
\usepackage{enumitem}
\usepackage{tikz}
\usepackage{graphicx}
\usepackage{csquotes}
\usepackage[acronym]{glossaries}
\usepackage{units}
\hypersetup{hidelinks}

\usepackage{bibgerm}
\usepackage[backend=bibtex,giveninits=true,maxnames=99,doi=false,isbn=false]{biblatex}
\tikzset{>=stealth}
\usetikzlibrary{arrows,shapes,calc}
\usetikzlibrary{positioning}
\usetikzlibrary{decorations}

\tikzset{%
        mid arrow/.style={postaction={decorate,decoration={				markings,mark=at position .5 with {\arrow[#1]{latex}}}}},
        mid arrow reversed/.style={postaction={decorate,decoration={				markings,mark=at position .5 with {\arrowreversed[#1]{latex}}}}},
		block/.style    = {draw, thick, rectangle, minimum height = 3em,
			minimum width = 2.5em, node distance = 6em, colorODE},
		dots/.style    = {draw = none},
		PDE/.style    = {draw, thick, rectangle, minimum height = 3em,
			minimum width = 8em, node distance = 6em, colorPDE},
		sum/.style      = {draw, circle, node distance = 2cm},
		input/.style    = {coordinate},
		output/.style   = {coordinate},
		gain/.style     = {draw, isosceles triangle, anchor = west,shape border rotate =-90, inner sep = 1pt, minimum size = 3em},
        box/.style    = {draw, thick, rectangle, minimum height = 3em,
			minimum width = 6.9em, text width = 6em, align = center, node distance = 6em, color3, fill=color3!5}
}

\definecolor{colorODE}{RGB}{0,167,253}
\definecolor{colorPDE}{RGB}{253,127,0}
\definecolor{colorInput}{RGB}{253,0,253}
\definecolor{color1}{RGB}{222,64,166}
\definecolor{color2}{RGB}{54,159,54}
\definecolor{color3}{RGB}{0,0,200}

\newcommand{\vect}[1]{\bm{#1}}
\newcommand{\mat}[1]{\vect{#1}}
\newcommand{\tint}{\int}
\newcommand{\rme}{\text{e}}
\newcommand{\Rset}{\mathbb R}
\newcommand{\Nset}{\mathbb N}
\newcommand{\dd}{\mathrm{d}}
\newcommand{\soll}{\mathrm{r}}
\DeclareMathOperator{\re}{Re}
\newcommand{\ubar}[1]{\text{\b{$#1$}}}

\newtheorem{rem}{Remark}
\newtheorem{example}{Example}
\theoremstyle{dgdef}\newtheorem{lem}{Lemma}
\theoremstyle{dgdef}\newtheorem{thm}{Theorem}

\graphicspath{{figs/}{logos/}}
\newcommand{\myvspace}[1]{\vspace{#1}}

\global\hbadness=10000

\glsdisablehyper 
\newacronym[shortplural={BCs}, longplural={boundary conditions}]{BC}{BC}{boundary condition}
\newacronym[shortplural={ICs}, longplural={initial conditions}]{IC}{IC}{initial condition}
\newacronym[shortplural={HCCFs}, longplural={hyperbolic controller canonical forms}]{HCCF}{HCCF}{hyperbolic controller canonical form}
\newacronym[shortplural={CCFs}, longplural={controller canonical forms}]{CCF}{CCF}{controller canonical form}
\newacronym[shortplural={OCFs}, longplural={observer canonical forms}]{OCF}{OCF}{observer canonical form}
\newacronym[shortplural={DPSs}, longplural={distributed-parameter systems}]{DPS}{DPS}{distributed-parameter system}
\newacronym{SISO}{SISO}{single-input single-output}

\begin{document}

  \articletype{Survey}

  \author*[1]{Nicole Gehring}
  \author[2]{Abdurrahman Irscheid}
  \author[3]{Joachim Deutscher}
  \author[4]{Frank Woittennek}
  \author[2]{Joachim Rudolph}
  \runningauthor{N. Gehring et al.}
  
  \affil[1]{Johannes Kepler University Linz, Austria}
  \affil[2]{Saarland University, Germany}
  \affil[3]{Ulm University, Germany}
  \affil[4]{UMIT TIROL, Austria}
  
  \title{Control of distributed-parameter systems using normal forms: An introduction}
  \runningtitle{Control of distributed-parameter systems using normal forms}

\abstract{This paper gives an overview of the control of distributed-parameter systems using normal forms.
Considering linear controllable PDE-ODE systems of hyperbolic type, two methods derive tracking controllers by mapping the system into a form that is advantageous for the control design, analogous to the finite-dimensional case. 
A flatness-based controller makes use of the hyperbolic controller canonical form that follows from a parametrization of the system's solutions.
A backstepping design exploits the strict-feedback form of the system to recursively stabilize and transform the subsystems.}

  \keywords{distributed-parameter systems, state feedback, backstepping, flatness, canonical forms}
  
%
  

\maketitle

  \section{Introduction}

\begin{figure}
    \centering
    \includegraphics[width=\columnwidth]{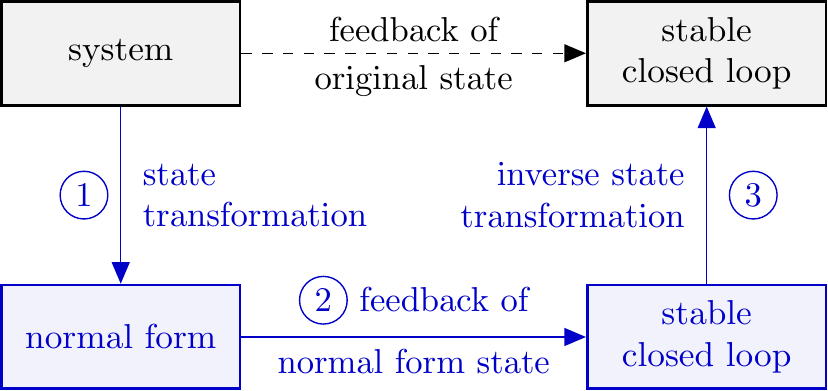}
    \caption{Design of a stabilizing feedback of the original state based on the three steps involving a normal form.}
    \label{fig:normalform-overview}
\end{figure}

In control theory, a normal or canonical form usually refers to a special state-space representation of a dynamical system that is well-suited for a certain analysis or synthesis purpose.
Thereby, the label canonical is usually used if a form is unique.
The basic idea of a controller design using normal forms is sketched in Figure~\ref{fig:normalform-overview}.
For linear finite-dimensional \gls{SISO} systems, the pioneering work \cite{Kalman1963siam} introduced the \gls{CCF} and the \gls{OCF} that are also sometimes called the controllable and observable canonical form, respectively.
While any observable system can be mapped into \gls{OCF}, the \gls{CCF}
\begin{equation}
\label{eq:ccf_finite}
    \dot{\vect\eta} = \begin{bmatrix}
        0 & 1 & 0 & \cdots & 0 \\
        \vdots & \ddots & \ddots & \ddots & \vdots \\
        \vdots & & \ddots & \ddots & 0 \\
        0 & \cdots & \cdots & 0 & 1 \\
        -a_0 & -a_1 & \cdots & \cdots & -a_{n-1}
    \end{bmatrix} \vect\eta + \begin{bmatrix} 0  \\ \vdots \\ \vdots \\ 0 \\ 1\end{bmatrix} u
\end{equation}
exists for any controllable system.
It is well known that the design of a stabilizing state feedback for \eqref{eq:ccf_finite} is straightforward as the characteristic coefficients $a_i$, $i=0,\dots,n-1$, are simply replaced by some desired ones of a stable closed-loop system (e.g.\ \cite{Rugh1996}).
This simplicity motivates the use of normal forms for a design.
As illustrated in Figure~\ref{fig:normalform-overview}, instead of a potentially very involved synthesis based on the original representation, a system is first mapped into, e.g., \gls{CCF}, with a stabilizing feedback of the normal form state implied by the obtained simple system structure.
Together with the inverse state transformation this results in the sought feedback of the original state. 
This design strategy using normal forms is not limited to linear finite-dimensional systems.

The nonlinear counterpart of \eqref{eq:ccf_finite} also involves a chain of integrators, with a nonlinear function in the last differential equation (e.g.~\cite{Zeitz1989ncs}).
As in the linear setting, the existence of such a nonlinear \gls{CCF} is closely related to the controllability of the system.
Flatness or more precisely differential flatness (see, e.g., \cite{Fliess1995ijc,Levine2009,Rudolph2021}) is a system property that can be interpreted as one possible extension of the notion of controllability to nonlinear systems.
If a system is flat, there exists a (fictitious) flat output that allows to express all system variables -- in particular the state and the input -- in terms of the flat output and a finite number of its time derivatives.
Not only is this flatness-based parametrization known to allow for an easy trajectory planning and feedforward design, but the flat output and its derivatives also form a nonlinear \gls{CCF} state.
Consequently, the controller synthesis for flat systems is straightforward.
For example, all nonlinear \gls{SISO} systems in strict-feedback form
\begin{subequations}
\label{eq:strictfeedback}
    \begin{align}
        \dot \eta_1 &= f_1(\eta_1) + g_1(\eta_1)\eta_2 \\
        \dot \eta_2 &= f_2(\eta_1,\eta_2) + g_2(\eta_1,\eta_2)\eta_3 \\
        &\,\,\vdots \nonumber \\
        \dot \eta_n &= f_n(\eta_1,\dots,\eta_n) + g_n(\eta_1,\dots,\eta_n)u
    \end{align}
\end{subequations}
(see, e.g., \cite{Krstic1995}) are flat due to their structure, with the differential equation for $\eta_i(t)$, $i=1,\dots,n$, only depending on $\eta_1(t),\dots,\eta_{i+1}(t)$ (with $\eta_{n+1}(t)=u(t)$).
Thus, $\eta_1(t)$ constitutes a flat output.
The well-known backstepping method (see, e.g., \cite{Krstic1995}) takes advantage of this recursive system structure in order to successively stabilize the $n$ subsystems by choice of the virtual control input $\eta_{i+1}(t)$ in the $i$-th design step.
As the virtual feedback induces a state transformation for the next step, with the actual state feedback in the final step, this too can be interpreted as an approach using normal forms (cf.~Figure~\ref{fig:normalform-overview}).

Importantly, both the backstepping design and the flatness-based approach have counterparts in the infinite-dimensional setting.
By applying the recursive backstepping design to a spatially discretized (infinite-dimensional) reaction-diffusion equation, it is shown in \cite{Balogh2002EJC} that a special choice of the virtual feedbacks corresponds to a Volterra integral transformation for the original \gls{DPS}.
Again, the idea is to map a given system into a form, from which a stabilizing feedback is easily deduced.
Nowadays, backstepping for parabolic and hyperbolic \glspl{DPS} (see, e.g., \cite{Krstic2008,Meurer2012}) is oftentimes used synonymous for a transformation-based design of controllers, with transformations not limited to those of Volterra type but also of Fredholm type or even more involved ones.
In contrast to this well-established and highly popular method, the flatness-based design in \cite{Woittennek2013CPDE,Woittennek2012at} is lesser known.
It builds on the works \cite{Russell1978SIAM} and \cite{Russell1978JMAA} that consider the controllability of linear hyperbolic \glspl{DPS}  and generalize the \gls{CCF} to the respective \gls{HCCF}.
The \gls{HCCF} in \cite{Russell1978JMAA} is introduced based on an equivalent representation of the hyperbolic system as a neutral differential equation, i.e., one with delays occurring in the highest time derivative (see \cite{Fridman2014}).
This delay representation directly follows from a flatness-based parametrization (see \cite{Woittennek2012mathmod}).
In the end, similar to the \gls{CCF} \eqref{eq:ccf_finite}, the control design based on the \gls{HCCF} only requires replacing the systems' characteristic with a desired one for the closed loop.

This paper gives an introduction to the systematic design of tracking controllers for \glspl{DPS} using normal forms.
In order to have a common set-up for both the flatness-based and the backstepping approach, here, the focus is on linear heterodirectional hyperbolic systems, where the partial differential equations (PDEs) are bidirectionally coupled with ordinary differential equations (ODEs) at one boundary and actuated at the other one.
Due to the interconnection between an ODE and a PDE subsystem, these \glspl{DPS} are also referred to as PDE-ODE systems.
They arise, for example, from technical processes like the axial and torsional vibrations of drilling systems, heavy ropes with a load as well as networks of open channels and transmission lines (see, e.g., \cite{Bastin2016, Petit2001siam, Gehring2022letters}).
Backstepping controllers for hyperbolic PDE-ODE systems are derived in many works, including \cite{Zhou2012ijc,Deutscher2017WC,DiMeglio2018aut,Auriol2018AUT,Deutscher2019ijc,Auriol2021ecc}.
Here, the multi-step design suggested in \cite{Deutscher2017WC} is used that exploits the strict-feedback form of the PDE-ODE system for a recursive stabilization (see also \cite{Gehring2021MTNS,Auriol2021ecc}).
The flatness-based approach essentially corresponds to \cite{Woittennek2013CPDE} and is presented using a constructive and illustrative perspective.
Ultimately, a survey of the control design for \glspl{DPS} using normal forms is given in this paper, with a focus on an introductory presentation and leaving in-depth mathematical background to cited references.

In what follows, Section~\ref{sec:problem} introduces the considered system class and specifies the control objective.
Based on two preliminary transformations, the system is mapped into a simpler form for the subsequent design of two tracking controllers.
First, the flatness-based design in Section~\ref{sec:flatness} makes use of a parametrization of the system's solutions in order to obtain a feedforward controller and the \gls{HCCF}, which in turn yields the flatness-based tracking controller.
The backstepping method in Section~\ref{sec:backstepping} capitalizes on the strict-feedback form of the system to successively stabilize it.
While both approaches are discussed independent of one another, Section~\ref{sec:comparison} goes into commonalities and differences.
The example of a heavy rope with a load in Section~\ref{sec:example} serves to illustrate the results and to give insight into the implementation of both controllers.
Finally, some details on existing as well as possible extensions are presented in Section~\ref{sec:conclusion}.

\textit{Notation:} The elements of a vector $\vect v\in\Rset^n$ are denoted by $v_i=\vect e_i^\top\vect v$, $i=1,\dots,n$, with the canonical unit vectors $\vect e_i\in\Rset^n$.
Similarly, double indices indicate elements $M_{ij}$ of a matrix $\mat M=[M_{ij}]$.
The identity matrix is denoted by $\mat I$.
As always, $\Rset$ and $\Nset$ are the set of real and natural numbers, respectively, with the latter including zero.
The notation $f\in C^n(\Omega)$ means that a function $f:\Omega\to\Rset$ is $n$ times piecewise continuously differentiable, $f\in L^2(\Omega)$ that it is absolutely square integrable.

  \section{Problem statement}
\label{sec:problem}

The class of linear hyperbolic PDE-ODE systems under consideration is visualized in Figure~\ref{fig:system_structure} by their coupling structure.
Preliminary transformations map the system into a simplified representation that is advantageous for the design of tracking controllers in Sections~\ref{sec:flatness} and \ref{sec:backstepping}.

\subsection{System class}

The central element of the system is a second-order heterodirectional hyperbolic PDE given in terms of the two coupled transport equations
\begin{subequations}
\label{eq:sys_pre}
    \begin{align}
    \label{eq:sys_pre_pde1}
        \partial_t x_1(z,t) &= \lambda_1(z) \partial_z x_1(z,t) + A_{11}(z) x_1(z,t) \nonumber \\
        &\hspace{1.5cm} + A_{12}(z) x_2(z,t), \quad z\in[0,1) \\
    \label{eq:sys_pre_pde2}
        \partial_t x_2(z,t) &= -\lambda_2(z) \partial_z x_2(z,t) + A_{21}(z) x_1(z,t) \nonumber \\
        &\hspace{1.5cm} + A_{22}(z) x_2(z,t), \quad z\in(0,1].
    \end{align}
    Assuming $\lambda_1(z),\lambda_2(z)>0$, the component $x_1(z,t)=\vect e_1^\top\vect x(z,t)$ of the PDE state $\vect x(z,t)\in\Rset^2$ propagates in the negative direction of the normalized spatial domain $[0,1]$ and $x_2(z,t)=\vect e_2^\top\vect x(z,t)$ in the positive $z$-direction, from $0$ to $1$.
    The transport velocities $\lambda_1$ and $\lambda_2$ are assumed to be piecewise continuously differentiable, with $a_{ij}\in C([0,1])$ for the four coupling functions.
    Due to a dynamic boundary condition at $z=0$, the transport equations \eqref{eq:sys_pre_pde1} and \eqref{eq:sys_pre_pde2} are bidirectionally coupled with the ODE
    \begin{equation}
    \label{eq:sys_pre_ode}
        \dot{\vect\xi}(t) = \mat F\vect\xi(t) + \vect b x_1(0,t)
    \end{equation}
    by means of the $x_1(0,t)$ acting on \eqref{eq:sys_pre_ode} and the ODE state $\vect\xi(t)\in\Rset^n$ affecting the \gls{BC}
    \begin{equation}
    \label{eq:sys_pre_bc0}
        x_2(0,t) = q_0 x_1(0,t) + \vect c^\top \vect\xi(t)
    \end{equation}
    of \eqref{eq:sys_pre_pde2}.
    The dimensions of $\mat F$, $\vect b$ and $\vect c^\top$ follow from that of $\vect\xi(t)$.
    The remaining \gls{BC}
    \begin{equation}
    \label{eq:sys_pre_bc1}
        x_1(1,t) = q_1 x_2(1,t) + u(t)
    \end{equation}
\end{subequations}
of \eqref{eq:sys_pre_pde1} introduces the control input $u(t)\in\Rset$.

\begin{figure}
    \centering
    \includegraphics[width=\columnwidth]{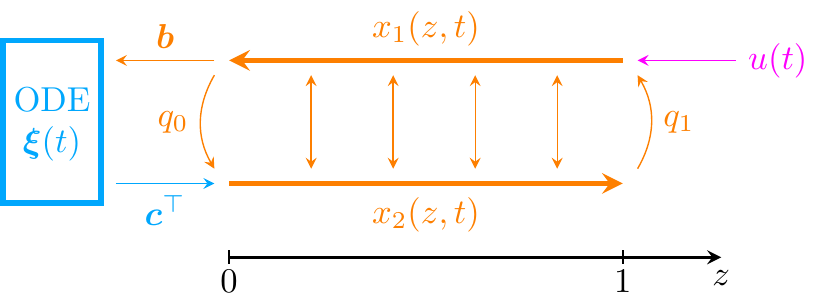}
    \caption{Visualization of the PDE-ODE system \eqref{eq:sys}.}
    \label{fig:system_structure}
\end{figure}

Using the matrix
\begin{equation}
    \mat\Lambda(z) = \begin{bmatrix} \lambda_1(z) & 0 \\[.5ex] 0 & -\lambda_2(z) \end{bmatrix}
\end{equation}
that comprises the transport velocities and $\mat A(z)=[A_{ij}(z)]$, \eqref{eq:sys_pre_pde1}--\eqref{eq:sys_pre_bc1} can be written in the compact form
\begin{subequations}
\label{eq:sys}
    \begin{align}
    \label{eq:sys_ode}
        \dot{\vect\xi}(t) &= \mat F\vect\xi(t) + \vect b x_1(0,t) \\
    \label{eq:sys_bc0}
        x_2(0,t) &= q_0 x_1(0,t) + \vect c^\top \vect\xi(t) \\
    \label{eq:sys_pde}
        \partial_t \vect x(z,t) &= \mat\Lambda(z) \partial_z \vect x(z,t) + \mat A(z) \vect x(z,t) \\
    \label{eq:sys_bc1}
        x_1(1,t) &= q_1 x_2(1,t) + u(t).
    \end{align}
\end{subequations}
The equations in \eqref{eq:sys} are arranged from $z=0$ to $z=1$ to reflect the strict-feedback form of the system, with $\vect\xi(t)$ essentially taking the role of $\eta_1(t)$ in \eqref{eq:strictfeedback} and $\vect x(z,t)$ that of $\eta_2(t)$.
Specifically, the ODE subsystem \eqref{eq:sys_ode} is driven by the PDE state at $z=0$ only.
In turn, the PDE subsystem \eqref{eq:sys_bc0}--\eqref{eq:sys_bc1} involves the ODE and the PDE state as well as the input.
Due to the bidirectional coupling between both subsystems, \eqref{eq:sys} is a PDE-ODE system.
Figure~\ref{fig:system_structure} illustrates the system structure, with dedicated colors for the ODE and PDE subsystem throughout the paper that also highlight the interconnection of both subsystems via $\vect b$ and $\vect c^\top$.
Introduce positive-definite functions
\begin{equation}
\label{eq:transporttimes}
    \phi_i(z) = \tint_0^z \frac{\dd\zeta}{\lambda_i(\zeta)}, \qquad z\in[0,1],\,i=1,2
\end{equation}
with the inverses $\psi_i(z)$ satisfying $\psi_i(\phi_i(z))=z$ as well as delays $\tau_i=\phi_i(1)$.
Then it is also apparent from Figure~\ref{fig:system_structure} that the input $u(t)$ at $z=1$ acts on the ODE at $z=0$ only after a time delay $\tau_1$ induced by the finite speed $\lambda_1(z)$ of propagation.
Given appropriate \glspl{IC} $\vect\xi(0)=\vect\xi_0\in\Rset^n$ and $\vect x(z,0)=\vect x_0(z)\in\Rset^2$, with piecewise continuous $\vect x_0$, the system \eqref{eq:sys} with $t>0$ is well posed\footnote{This paper does not discuss the abstract notion of state spaces. Still, a typical choice for \eqref{eq:sys} with $\vect\xi(t)$ and $\vect x(z,t)$ would be $\Rset^n\times (L^2([0,1]))^2$.} (e.g.\ \cite[App.~A]{Bastin2016}).

\myvspace{.75cm}
\subsection{Control objective}

The control objective is to make the states $\vect x(z,t)$ and $\vect\xi(t)$ converge to the corresponding, predefined reference trajectories in
\begin{equation}
\label{eq:sys_reference}
    (\vect x_\soll(z,t),\vect\xi_\soll(t),u_\soll(t)).
\end{equation}
While the specification of \eqref{eq:sys_reference} is discussed in the context of Section~\ref{sec:flatness_tracking}, in general, the reference is only assumed to satisfy the dynamics \eqref{eq:sys}.
The following two assumptions are imposed for the design of a (static) state-feedback tracking controller for the system \eqref{eq:sys} with input $u(t)$:
\begin{enumerate}[label=(A\arabic*), leftmargin=1cm]

  \item \label{enum:ass_ode} The pair $(\mat F,\vect b)$ is controllable.
  
  \item \label{enum:ass_pde} In the \gls{BC} \eqref{eq:sys_bc0}, $q_0\neq0$ holds.
  
\end{enumerate}
Roughly speaking, \ref{enum:ass_ode} ensures the controllability of the ODE subsystem and \ref{enum:ass_pde} of the PDE subsystem.
Noting that the boundary value $x_1(0,t)$ takes the role of an input w.r.t.\ the ODE \eqref{eq:sys_ode}, the necessity of \ref{enum:ass_ode} is evident from classical theory of linear finite-dimensional systems (e.g.\ \cite[Thm.~9.5]{Rugh1996}).
Assumption~\ref{enum:ass_pde} guarantees exact controllability of the PDE subsystem (e.g.~\cite[Thm.~3.2]{Russell1978SIAM}).
This too is necessary as $q_0=0$ only gives null controllability, a weaker property that is closely related to stabilizability and merely allows to steer the system to zero\footnote{Stabilization 
is still possible if \ref{enum:ass_pde} is not met and even if $(\mat F,\vect b)$ is only stabilizable (e.g.\ \cite{Deutscher2017WC,DiMeglio2018aut,Deutscher2019ijc}). The choice of closed-loop dynamics is more restricted in this case.}.
In contrast, exact controllability means that all states can be reached from the origin.
It can also be interpreted as null controllability in forward and backward time.
In the end, \ref{enum:ass_ode} and \ref{enum:ass_pde} guarantee  controllability of the PDE-ODE system \eqref{eq:sys}, an obvious prerequisite for the existence of an infinite-dimensional \gls{CCF}.


\subsection{Preliminary transformations}
\label{sec:prelim}

Two preliminary transformations are invoked to map the system \eqref{eq:sys} into a form that is advantageous for the controller designs in Sections~\ref{sec:flatness} and \ref{sec:backstepping}.
First, the transformation
\begin{equation}
\label{eq:pre_trafo_hopfcole}
    \tilde{\vect x}(z,t) = \mat E(z)\vect x(z,t) = \begin{bmatrix} 
        \rme^{\alpha_1(z)} & 0 \\
        0 & \rme^{-\alpha_2(z)} \end{bmatrix} \vect x(z,t)
\end{equation}
with $\alpha_i(z)=\int_0^z \frac{A_{ii}(\zeta)}{\lambda_i(\zeta)}\,\dd\zeta$, $i=1,2$, allows to rewrite \eqref{eq:sys_pde} without the source terms coefficients $A_{ii}(z)$ (e.g.\ \cite{Coron2012SIAM}), thus simplifying the system structure.
Applying the scaling \eqref{eq:pre_trafo_hopfcole} to \eqref{eq:sys} yields
\begin{subequations}
\label{eq:sys_antidiag}
    \begin{align}
        \label{eq:sys_antidiag_ode}
        \dot{\vect\xi}(t) &= \mat F\vect\xi(t) + \vect b \tilde x_1(0,t) \\
        \label{eq:sys_antidiag_bc0}
        \tilde x_2(0,t) &= q_0 \tilde x_1(0,t) + \vect c^\top \vect\xi(t) \\
        \label{eq:sys_antidiag_pde}
        \partial_t \tilde{\vect x}(z,t) &= \mat\Lambda(z) \partial_z \tilde{\vect x}(z,t) + \tilde{\mat A}(z)\tilde{\vect x}(z,t) \\
        \label{eq:sys_antidiag_bc1}
        \tilde x_1(1,t) &= q_1\rme^{\alpha_1(1)+\alpha_2(1)}\tilde x_2(1,t) + \rme^{\alpha_1(1)}u(t).
    \end{align}
\end{subequations}
Therein, as a consequence of the transformation \eqref{eq:pre_trafo_hopfcole}, the in-domain coupling matrix
\begin{equation}
\label{eq:pre_Atilde}
    \tilde{\mat A}(z) = 
    \begin{bmatrix} 0 & \tilde A_{12}(z) \\[.5ex] \tilde A_{21}(z) & 0 \end{bmatrix}
\end{equation}
has zero diagonal elements, with the remaining entries defined by $\tilde A_{12}(z)=A_{12}(z)\rme^{\alpha_1(z)+\alpha_2(z)}$ and $\tilde A_{21}(z)=A_{21}(z)\rme^{-\alpha_1(z)-\alpha_2(z)}$.

\begin{figure}
    \centering
    \includegraphics{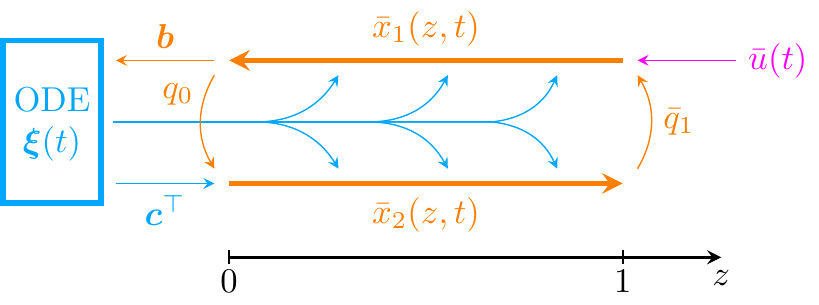}
    \caption{Visualization of the transformed PDE-ODE system \eqref{eq:sys_transp}.}
    \label{fig:systransp_structure}
\end{figure}

In the second preliminary step, a Volterra integral transformation
\begin{equation}
\label{eq:pre_trafo_bs}
    \bar{\vect x}(z,t) = \tilde{\vect x}(z,t) - \tint_0^z \mat K(z,\zeta)\tilde{\vect x}(\zeta,t)\,\dd\zeta
\end{equation}
is used to remove the remaining in-domain coupling attributed to $\tilde{\mat A}(z)$ in \eqref{eq:pre_Atilde} altogether, thus further simplifying the system representation.
For that, the kernel $\mat K(z,\zeta)\in\Rset^{2\times2}$, defined on the triangular domain
\begin{equation}
\label{eq:triangulardomein}
    \mathcal T = \{(z,\zeta)\in[0,1]^2 | \zeta\le z\},
\end{equation}
has to be chosen such that \eqref{eq:pre_trafo_bs} maps \eqref{eq:sys_antidiag} into the form
\begin{subequations}
\label{eq:sys_transp}
    \begin{align}
    \label{eq:sys_transp_ode}
        \dot{\vect\xi}(t) &= \mat F\vect\xi(t) + \vect b \bar x_1(0,t) \\
    \label{eq:sys_transp_bc0}
        \bar x_2(0,t) &= q_0 \bar x_1(0,t) + \vect c^\top \vect\xi(t) \\
    \label{eq:sys_transp_pde}
        \partial_t \bar{\vect x}(z,t) &= \mat\Lambda(z) \partial_z \bar{\vect x}(z,t) + \mat C(z)\vect\xi(t) \\
    \label{eq:sys_transp_bc1}
        \bar x_1(1,t) &= \bar q_1\bar x_2(1,t) + \bar u(t),
    \end{align}
\end{subequations}
where the matrix $\mat C(z)$ and the boundary value $\bar u(t)$ are disregarded for the moment.
It is apparent from Figure~\ref{fig:systransp_structure} that the structure of \eqref{eq:sys_transp} is less complex than that of \eqref{eq:sys} (see also Figure~\ref{fig:system_structure}), as the PDE subsystem essentially only comprises two cascaded transport equations, without any couplings.


In order to determine $\mat K(z,\zeta)$, the integral transformation \eqref{eq:pre_trafo_bs} is inserted in \eqref{eq:sys_transp} while keeping in mind that $\tilde{\vect x}(z,t)$ satisfies \eqref{eq:sys_antidiag}.
As $\bar{\vect x}(0,t)=\tilde{\vect x}(0,t)$, \eqref{eq:sys_transp_ode} and \eqref{eq:sys_transp_bc0} follow from \eqref{eq:sys_antidiag_ode} and \eqref{eq:sys_antidiag_bc0} without any conditions being imposed on $\mat K(z,\zeta)$.
Similarly, a comparison of \eqref{eq:sys_antidiag_bc1} and \eqref{eq:sys_transp_bc1} yields $\bar q_1=q_1\rme^{\alpha_1(1)+\alpha_2(1)}$ as well as the new input
\begin{equation}
\label{eq:input_trafo}
    \bar u(t) = \rme^{\alpha_1(1)}u(t) + \tint_0^1 (\bar q_1\vect e_2^\top-\vect e_1^\top) \mat K(1,z)\tilde{\vect x}(z,t)\,\dd z
\end{equation}
that is introduced for convenience of notation.
Note that $\bar u(t)$ is intentionally introduced such that the boundary term $\bar q_1\bar x_2(1,t)$ in \eqref{eq:sys_transp_bc1} is retained.
By that, both \eqref{eq:sys} and \eqref{eq:sys_transp} can be interpreted as time-delay systems of neutral type if $q_1\neq0$ (e.g.\ \cite{Fridman2014}), while \eqref{eq:sys_transp} would never be neutral if $\bar u(t)$ had compensated $\bar q_1\bar x_2(1,t)$.

\begin{figure}[t]
    \centering
    \includegraphics{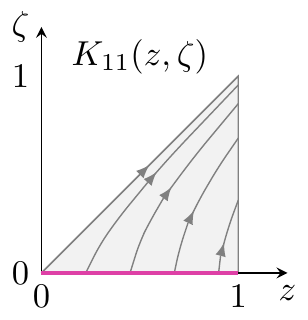} \quad
    \includegraphics{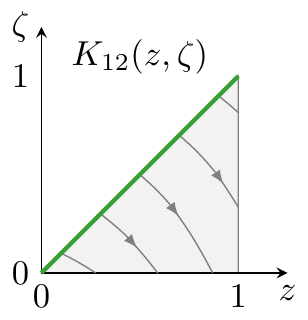} \\
    \includegraphics{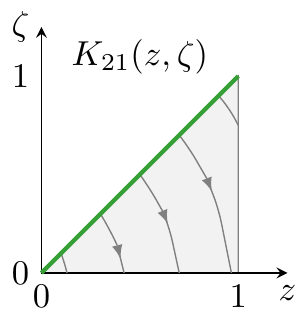} \quad
    \includegraphics{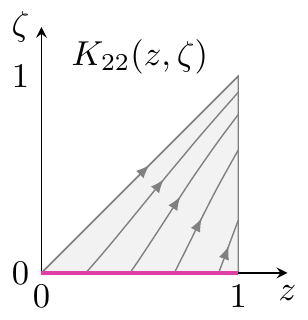}
    \caption{Typical characteristic curves visualize the kernel equations \eqref{eq:pre_kerneleqs}, with \glspl{BC} \eqref{eq:pre_kerneleqs_11} and \eqref{eq:pre_kerneleqs_22} at $\zeta=0$ highlighted in magenta, and \glspl{BC} \eqref{eq:pre_kerneleqs_12} and \eqref{eq:pre_kerneleqs_21} on the diagonal $\zeta=z$ in green.}
    \label{fig:kernel}
\end{figure}

Differentiating \eqref{eq:pre_trafo_bs} w.r.t.\ $t$ and substituting $\partial_t\tilde{\vect x}(z,t)$ using \eqref{eq:sys_antidiag_pde} as well as $\partial_z\tilde{\vect x}(z,t)$ by means of \eqref{eq:pre_trafo_bs}, an integration by part yields
\begin{align}
\label{eq:pre_pde_bilanz}
    &\partial_t \bar{\vect x}(z,t) = \mat\Lambda(z)\partial_z\bar{\vect x}(z,t) + \mat K(z,0)\mat\Lambda(0)\vect e_2\vect c^\top\vect\xi(t) \nonumber\\
    &\qquad+ \tint_0^z\big[\mat\Lambda(z)\partial_z\mat K(z,\zeta)+\partial_\zeta\big(\mat K(z,\zeta)\mat\Lambda(\zeta)\big) \nonumber\\
    &\hspace{4cm}-\mat K(z,\zeta)\tilde{\mat A}(\zeta)\big]\tilde{\vect x}(\zeta,t)\,\dd\zeta \nonumber\\
    &\qquad+ \left[\mat\Lambda(z)\mat K(z,z)-\mat K(z,z)\mat\Lambda(z)+\tilde{\mat A}(z)\right]\tilde{\vect x}(z,t) \nonumber\\
    &\qquad+ \left[\mat K(z,0)\mat\Lambda(0)(\vect e_1+q_0\vect e_2)\right]\tilde x_1(0,t)
\end{align}
in view of \eqref{eq:sys_antidiag_bc0}.
For \eqref{eq:pre_pde_bilanz} to equal \eqref{eq:sys_transp_pde} for arbitrary values of $\tilde{\vect x}(z,t)$, the matrix
\begin{equation}
    \mat C(z) = \mat K(z,0)\mat\Lambda(0)\vect e_2\vect c^\top
\end{equation}
in \eqref{eq:sys_transp_pde} is defined based on $\mat K(z,\zeta)$ and the terms in the square brackets in \eqref{eq:pre_pde_bilanz} have to vanish.
The latter results in the so-called kernel equations that comprise essentially four transport equations in
\begin{subequations}
\label{eq:pre_kerneleqs}
    \begin{align}
        \mat\Lambda(z)\partial_z \mat K(z,\zeta) + \partial_\zeta\big(\mat K(z,\zeta)\mat\Lambda(\zeta)\big) &= \mat K(z,\zeta)\tilde{\mat A}(\zeta)
    \end{align}
    for the elements $K_{ij}(z,\zeta)$ of $\mat K(z,\zeta)$ on the triangular domain $\mathcal T$ as well as the corresponding \glspl{BC}
    \begin{align}
    \label{eq:pre_kerneleqs_11}
        K_{11}(z,0) &= \frac{q_0\lambda_2(0)}{\lambda_1(0)}K_{12}(z,0) \\
    \label{eq:pre_kerneleqs_12}
        K_{12}(z,z) &= -\frac{\tilde A_{12}(z)}{\lambda_1(z)+\lambda_2(z)} \\
    \label{eq:pre_kerneleqs_21}
        K_{21}(z,z) &= \frac{\tilde A_{21}(z)}{\lambda_1(z)+\lambda_2(z)} \\
    \label{eq:pre_kerneleqs_22}
        K_{22}(z,0) &= \frac{\lambda_1(0)}{q_0\lambda_2(0)}K_{21}(z,0).
    \end{align}
\end{subequations}
The transformation \eqref{eq:pre_trafo_bs} and the kernel equations \eqref{eq:pre_kerneleqs} are well-known in the context of backstepping control (see, e.g., \cite{Krstic2008}).
Based on $A_{ij}\in C([0,1])$ and $\lambda_i\in C^1([0,1])$, $i,j=1,2$, it is shown in \cite{Vazquez2011CDC,Coron2012SIAM} that \eqref{eq:pre_kerneleqs} admits a unique continuous solution. 
It can be determined using the method of characteristics and a successive approximation.
This seems quite intuitive based on Figure~\ref{fig:kernel}, which illustrates the evolution of the solution $\mat K(z,\zeta)$ along typical characteristic curves on the triangular domain $\mathcal T$, starting at the colorized boundaries specified by \eqref{eq:pre_kerneleqs_11}--\eqref{eq:pre_kerneleqs_22}.

\begin{rem}[General system class]
    Based on the Volterra integral transformation \eqref{eq:pre_trafo_bs} and in particular \eqref{eq:pre_pde_bilanz}, it is evident that the PDE \eqref{eq:sys_pde} can be generalized to
    \begin{align}
        & \partial_t \vect x(z,t) = \mat\Lambda(z) \partial_z \vect x(z,t) + \mat A(z) \vect x(z,t) \\
        &\quad + \tint_0^z \mat D_0(z,\zeta)\vect x(\zeta,t)\,\dd\zeta + \vect d_1(z) x_1(0,t) + \mat D_2(z) \vect\xi(t), \nonumber
    \end{align}
    with additional suitable integral, local and ODE terms at the actuated boundary \eqref{eq:sys_bc1}.
    Essentially, the system is only required to be in strict feedback form (see \cite{Gehring2021MTNS}).
\end{rem}

As the preliminary transformations\footnote{The two transformations could be combined into a single one. However, this would yield more complex kernel equations.} \eqref{eq:pre_trafo_hopfcole} and \eqref{eq:pre_trafo_bs} make use of Assumption~\ref{enum:ass_pde} (see division by $q_0$ in \eqref{eq:pre_kerneleqs_22}), \eqref{eq:sys_transp} exists for all PDE-ODE systems with an exactly controllable PDE.
Let $\vect\xi(0)=\vect\xi_0$ and $\bar{\vect x}(z,0)=\mat E(z)\vect x_0(z)-\int_0^z\mat K(z,\zeta)\mat E(\zeta)\vect x_0(\zeta)\,\dd\zeta$ be the \glspl{IC} of \eqref{eq:sys_transp} defined based on those of \eqref{eq:sys}.
Then, the following lemma holds.

\begin{lem}[Equivalence of \eqref{eq:sys} and \eqref{eq:sys_transp}]
\label{lem:equiv_sys_pre}
    With the input $\bar u(t)$ defined via \eqref{eq:input_trafo}, \eqref{eq:sys_transp} is an equivalent representation of every system \eqref{eq:sys} with input $u(t)$ that satisfies \ref{enum:ass_pde}.
\end{lem}

Equivalence results from the invertibility of the transformations \eqref{eq:pre_trafo_hopfcole} and \eqref{eq:pre_trafo_bs}, the first being obvious from the regularity of the scaling matrix $\mat E(z)$, $z\in[0,1]$.
The kernel $\mat K_\mathrm{I}(z,\zeta)\in\Rset^{2\times2}$ of the inverse map
\begin{equation}
\label{eq:pre_trafo_bs_inv}
    \tilde{\vect x}(z,t) = \bar{\vect x}(z,t) + \tint_0^z \mat K_\mathrm{I}(z,\zeta)\bar{\vect x}(\zeta,t)\,\dd\zeta
\end{equation}
of \eqref{eq:pre_trafo_bs} follows from the so-called reciprocity relation
\begin{equation}
\label{eq:reciprocity}
    \mat K_\mathrm{I}(z,\zeta) = \mat K(z,\zeta) + \tint_\zeta^z \mat K(z,\bar\zeta)\mat K_\mathrm{I}(\bar\zeta,\zeta)\,\dd\bar\zeta
\end{equation}
(e.g.~\cite{Deutscher2019ijc}).
For every $\zeta\le z\in[0,1]$, this matrix-valued Volterra integral equation admits a unique continuous solution $\mat K_\mathrm{I}(z,\zeta)$ (e.g.\ \cite[Thm.~3.11]{Linz1987}), as the elements of $\mat K(z,\zeta)$ are $C([0,1]^2)$ functions.

The representation \eqref{eq:sys_transp} of system \eqref{eq:sys} serves as a common basis for the controller designs in Sections~\ref{sec:flatness} and \ref{sec:backstepping}.
Based on the references \eqref{eq:sys_reference} for \eqref{eq:sys}, the input $\bar u(t)$ is determined such that the dynamics \eqref{eq:sys_transp} is stabilized along a corresponding reference trajectory $(\bar{\vect x}_\soll(z,t),\vect\xi_\soll(t),\bar u_\soll(t))$.
The function $\bar{\vect x}_\soll(z,t)$ 
follows from substituting \eqref{eq:sys_reference} into the transformations \eqref{eq:pre_trafo_hopfcole} and \eqref{eq:pre_trafo_bs}, with $\bar u_\soll(t)$ 
implied by \eqref{eq:input_trafo}, \eqref{eq:pre_trafo_hopfcole} and \eqref{eq:sys_reference}.
Lemma~\ref{lem:equiv_sys_pre} ensures that any tracking controller for \eqref{eq:sys_transp} also guarantees the stabilization of the corresponding reference for \eqref{eq:sys}.

  \section{Flatness-based design}
\label{sec:flatness}

The design in \cite{Woittennek2013CPDE} makes use of a special state-space representation of the PDE-ODE system \eqref{eq:sys_transp}, the \gls{HCCF}, that is well-suited for the control task.
The basis for the \gls{HCCF} is a flatness-based parametrization of the system variables $\bar{\vect x}(z,t)$, $\vect\xi(t)$ and $\bar u(t)$ in terms of a flat output $y(t)$.
In contrast to finite-dimensional systems, due to the hyperbolic nature of \eqref{eq:sys_transp}, parametrizing all system solutions involves not only derivatives of $y(t)$ but also time shifts.
The \gls{HCCF} of \eqref{eq:sys_transp} is implied by the flatness-based parametrization of the input $\bar u(t)$.
From that, the choice of a desired stable closed-loop dynamics directly gives the flatness-based controller.

\begin{figure*}[t]
    \centering
    \scalebox{.8}{\includegraphics{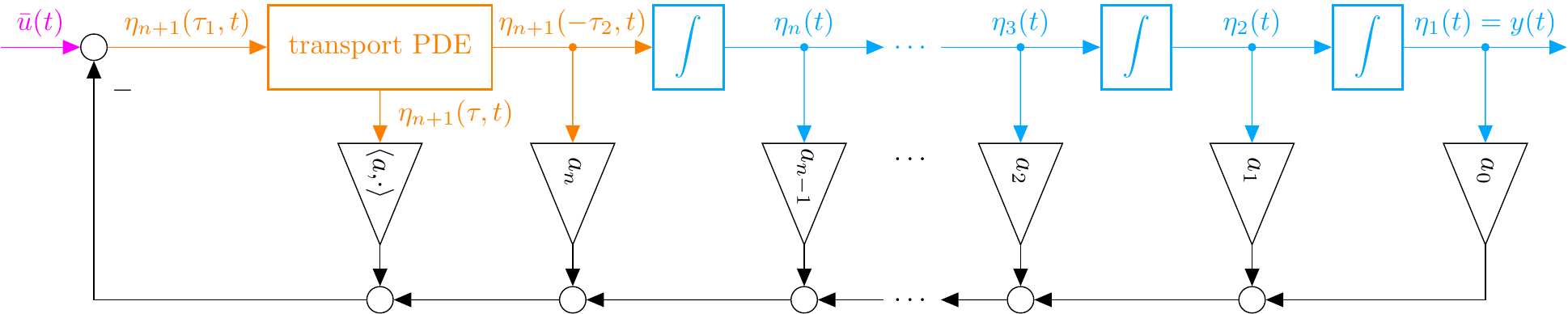}}
    \caption{Visualization of the \gls{HCCF} \eqref{eq:flat_hccf} by means of a signal flow diagram, where $\langle a,\eta_{n+1}(\cdot,t)\rangle=\int_{-\tau_2}^{\tau_1} a(\tau)\eta_{n+1}(\tau,t)\,\dd\tau$.}
    \label{fig:HCCF}
\end{figure*}

\subsection{Flatness-based parametrization}
\label{sec:flat_para}

It is well known that the ODE \eqref{eq:sys_transp_ode}, which is controllable by Assumption~\ref{enum:ass_ode}, can be transformed into a \gls{CCF} analogous to \eqref{eq:ccf_finite}.
Due to the chain of integrators in this special state-space representation, with the input acting only on the last state component, the first state component of the \gls{CCF} state is a flat output.
Using the (invertible) Kalman controllability matrix $\mat M_c=[\vect b,\mat F\vect b,\dots,\mat F^{n-1}\vect b]$, a flat output can be written as
\begin{equation}
\label{eq:flat_output}
    y(t) = \vect e_n^\top \mat M_c^{-1}\vect\xi(t) = \vect h^\top\vect\xi(t)
\end{equation}
(e.g.\ \cite{Rugh1996, Levine2009}). Denote by $\vect y^{[k]}(t)=[y(t),\dot y(t),\dots,y^{(k)}(t)]^\top$, $k\in\Nset$, the vector containing the flat output $y(t)$ and its derivatives up to order $k$, which is also the \gls{CCF} state for $k=n-1$.
Then, based on the state transformation
\begin{equation}
    \vect y^{[n-1]}(t) = \mat T_c \vect\xi(t) \quad\text{with}\quad \mat T_c = \begin{bmatrix} \vect h^\top \\ \vect h^\top\mat F \\ \vdots \\ \vect h^\top\mat F^{n-1} \end{bmatrix}
\end{equation}
(e.g.\ \cite{Rugh1996,Levine2009}), standard calculations confirm that \eqref{eq:flat_output} provides the differential parametrization
\begin{subequations}
\label{eq:flat_parametrization_z0}
    \begin{align}
    \label{eq:flat_parametrization_z0_xi}
        \vect\xi(t) &= \mat T_c^{-1} \vect y^{[n-1]}(t) \\
    \label{eq:flat_parametrization_z0_x10}
        \bar x_1(0,t) &= y^{(n)}(t) - \vect e_n^\top\mat T_c\mat F\mat T_c^{-1} \vect y^{[n-1]}(t) \nonumber \\
        &= \vect p_1^\top\vect y^{[n]}(t)
        \intertext{of the quantities in \eqref{eq:sys_transp_ode}, i.e., the ODE state $\vect\xi(t)$ and the boundary value $\bar x_1(0,t)$ are expressed in terms of $y(t)$ and its derivatives.
        Substituting these into the \gls{BC} \eqref{eq:sys_transp_bc0} gives}
    \label{eq:flat_parametrization_z0_x20}
        \bar x_2(0,t) &= q_0 \vect p_1^\top\vect y^{[n]}(t) + \vect c^T \mat T_c^{-1} \vect y^{[n-1]}(t) \nonumber \\
        &= \vect p_2^\top\vect y^{[n]}(t).
    \end{align}
\end{subequations}
By \eqref{eq:flat_parametrization_z0}, all boundary values at $z=0$ are parametrized.

In order to obtain a flatness-based parametrization of the distributed state $\bar{\vect x}(z,t)$, the PDE \eqref{eq:sys_transp_pde} is solved by integration along the characteristic curves.
This is significantly facilitated due to the preliminary transformation \eqref{eq:pre_trafo_bs}.
Starting at the boundary $z=0$, in view of \eqref{eq:transporttimes}, the solution of the Cauchy problem w.r.t.\ $z$ reads
\begin{subequations}
\label{eq:flat_sol_pde}
    \begin{align}
    \label{eq:flat_sol_pde1}
        \bar x_1(z,t) &= \bar x_1(0,t+\phi_1(z)) \\
        &\hspace{1cm} - \tint_0^z \frac{\vect e_1^\top\mat C(\zeta)}{\lambda_1(\zeta)} \vect\xi(t+\phi_1(z)-\phi_1(\zeta))\,\dd\zeta \nonumber \\
    \label{eq:flat_sol_pde2}
        \bar x_2(z,t) &= \bar x_2(0,t-\phi_2(z)) \\
        &\hspace{1cm} + \tint_0^z \frac{\vect e_2^\top\mat C(\zeta)}{\lambda_2(\zeta)} \vect\xi(t-\phi_2(z)+\phi_2(\zeta))\,\dd\zeta. \nonumber
    \end{align}
\end{subequations}
Then, simply substituting \eqref{eq:flat_parametrization_z0} into \eqref{eq:flat_sol_pde} yields the flatness-based parametrization
\begin{subequations}
\label{eq:flat_x_para}
    \begin{align}
    \label{eq:flat_x_para1}
        \bar x_1(z,t) &= \vect p_1^\top \vect y^{[n]}(t+\phi_1(z)) \\
        &\hspace{.3cm} - \tint_0^z \frac{\vect e_1^\top\mat C(\zeta)}{\lambda_1(\zeta)} \mat T_c^{-1} \vect y^{[n-1]}(t+\phi_1(z)-\phi_1(\zeta))\,\dd\zeta \nonumber \\
    \label{eq:flat_x_para2}
        \bar x_2(z,t) &= \vect p_2^\top \vect y^{[n]}(t-\phi_2(z)) \\
        &\hspace{.3cm} + \tint_0^z \frac{\vect e_2^\top\mat C(\zeta)}{\lambda_2(\zeta)} \mat T_c^{-1} \vect y^{[n-1]}(t-\phi_2(z)+\phi_2(\zeta))\,\dd\zeta \nonumber
    \end{align}
\end{subequations}
of the distributed state $\bar{\vect x}(z,t)$,
which involves $y(t)$ and its derivatives as well as time shifts thereof.
As $\phi_i(z)\ge\phi_i(\zeta)\ge0$, $i=1,2$, for all $\zeta\in[0,z]$, any time shift in \eqref{eq:flat_x_para1} is positive, thus corresponding to a prediction, with only negative time shifts, i.e.\ delays, occurring in \eqref{eq:flat_x_para2}.
The flatness-based parametrization of the system variables in \eqref{eq:sys_transp} is completed by expressing the input in terms of $y(t)$.
To this end, recalling that $\psi_i(\phi_i(z))=z$ and $\tau_i=\phi_i(1)$, $i=1,2$, \eqref{eq:flat_x_para} is used in the \gls{BC} \eqref{eq:sys_transp_bc1} at $z=1$ to give
\begin{align}
\label{eq:flat_baru_para}
    \bar u(t) &= \vect p_1^\top \vect y^{[n]}(t+\tau_1) - \bar q_1 \vect p_2^\top \vect y^{[n]}(t-\tau_2) \\
    &\hspace{.3cm} - \tint_0^{\tau_1} \vect e_1^\top\mat C(\psi_1(\tau_1-\tau)) \mat T_c^{-1} \vect y^{[n-1]}(t+\tau)\,\dd\tau \nonumber \\
    &\hspace{.3cm} - \tint_0^{\tau_2} \bar q_1\vect e_2^\top\mat C(\psi_2(\tau_2-\tau)) \mat T_c^{-1} \vect y^{[n-1]}(t-\tau)\,\dd\tau \nonumber
\end{align}
after a transformation of the integration variable, in order to better highlight the different time shifts.
Due to the time shifts in the highest-order derivative of $y(t)$ in the functional differential equation \eqref{eq:flat_baru_para}, the input parametrization also constitutes a neutral-type time-delay system.
Importantly, \eqref{eq:flat_baru_para} is equivalent to \eqref{eq:sys_transp} in that it is simply a different way of writing the dynamics (e.g.\ \cite{Russell1991}).

\myvspace{.5cm}
\subsection{Hyperbolic controller canonical form}

In accordance with \cite{Russell1991} and \cite{Woittennek2012mathmod}, the \gls{HCCF} is introduced based on the input parametrization \eqref{eq:flat_baru_para}.
For that, a thorough inspection of \eqref{eq:flat_baru_para} reveals that $\bar u(t)$ explicitly depends on $y^{(n)}(t+\tau_1)$, which is apparent from the definition of $\vect p_1^\top$ in \eqref{eq:flat_parametrization_z0_x10}.
Moreover, time shifts cover delays with a maximum amplitude of $\tau_2$ and predictions up to $\tau_1$.
This motivates the definition of the \gls{HCCF} state\footnote{A typical state space would be $\Rset^n\times L^2([-\tau_2,\tau_1])$. See \cite{Woittennek2013CPDE} for details on state spaces in the context of flatness-based control.}
\begin{subequations}
\label{eq:flat_hccf_state}
    \begin{align}
    \label{eq:flat_hccf_state_ode}
        \eta_i(t) &= y^{(i-1)}(t-\tau_2), && i = 1,\dots,n \\
    \label{eq:flat_hccf_state_pde}
        \eta_{n+1}(\tau,t) &= y^{(n)}(t+\tau), && \tau\in[-\tau_2,\tau_1],
    \end{align}
\end{subequations}
where the dimension of the finite-dimensional part $\vect\eta(t)=[\eta_1(t),\dots,\eta_n(t)]^\top$ corresponds to the highest-order derivative in \eqref{eq:flat_baru_para} as well as the dimension of the ODE state $\vect\xi(t)$ in \eqref{eq:sys_transp}.
Since $\tau_1$ and $\tau_2$ are the time delays induced by the transport velocities $\lambda_1(z)$ and $\lambda_2(z)$ in \eqref{eq:sys_transp}, the definition \eqref{eq:flat_hccf_state_pde} relates to the PDE state $\vect{\bar x}(z,t)$ where $z\in[0,1]$.
Moreover, the domain $[-\tau_2,\tau_1]$ of the distributed state $\eta_{n+1}(\tau,t)$ is on par with the time shifts in \eqref{eq:flat_baru_para}, i.e., $\tau$ can be interpreted as another time argument of $\eta_{n+1}(\tau,t)$.

In accordance with \cite{Woittennek2012mathmod}, the \gls{HCCF} for a system \eqref{eq:sys} satisfying Assumptions \ref{enum:ass_ode} and \ref{enum:ass_pde} is defined as
\begin{subequations}
\label{eq:flat_hccf}
    \begin{align}
    \label{eq:flat_hccf_etai}
        \dot\eta_i(t) &= \eta_{i+1}(t), \qquad i=1,\dots,n-1 \\
    \label{eq:flat_hccf_etan}
        \dot\eta_n(t) &= \eta_{n+1}(-\tau_2,t) \\
    \label{eq:flat_hccf_pde}
        \partial_t\eta_{n+1}(\tau,t) &= \partial_\tau\eta_{n+1}(\tau,t), \qquad \tau\in[-\tau_2,\tau_1) \\
    \label{eq:flat_hccf_bc}
        \eta_{n+1}(\tau_1,t) &= - \tint_{-\tau_2}^{\tau_1} a(\tau)\eta_{n+1}(\tau,t)\,\dd\tau - \sum_{i=1}^n a_{i-1}\eta_i(t) \nonumber \\
        &\qquad - a_n\eta_{n+1}(-\tau_2,t) + \bar u(t)
    \end{align}
\end{subequations}
and generalizes the well-known \gls{CCF} \eqref{eq:ccf_finite} for linear ODEs to linear hyperbolic PDE-ODE systems.
Therein, \eqref{eq:flat_hccf_etai}--\eqref{eq:flat_hccf_pde} directly follow from the definition \eqref{eq:flat_hccf_state} of the \gls{HCCF} state.
It is illustrated in Figure~\ref{fig:HCCF} that the \gls{HCCF} comprises a chain of $n$ integrators (see \eqref{eq:flat_hccf_etai}--\eqref{eq:flat_hccf_etan}) that is also typical for \glspl{CCF} in the finite-dimensional case.
The integrators are preceded by the transport PDE \eqref{eq:flat_hccf_pde}, the state $\eta_{n+1}(\tau,t)$ of which propagates with a normalized velocity from $\tau=\tau_1$ to the boundary at $\tau=-\tau_2$, where $\eta_{n+1}(-\tau_2,t)$ serves as the input of the last integrator \eqref{eq:flat_hccf_etan}.
The normalization of the velocity as well as the action of the input $\bar u(t)$ at the boundary $\tau=\tau_1$, analogously to the input vector of the \gls{CCF} \eqref{eq:ccf_finite}, justifies the classification of the \gls{HCCF} \eqref{eq:flat_hccf} as a normal form\footnote{It could be argued that \eqref{eq:flat_hccf} is not canonical and only constitutes a hyperbolic controller form (HCF), as the domain $[-\tau_2,\tau_1]$ of the distributed state is fixed only up to a scaling and an offset, with $[-1,1]$ in \cite{Russell1978JMAA} and \cite{Russell1991}. However, it is only possible to normalize either the transport velocity in \eqref{eq:flat_hccf_pde} or the domain.}.

The coefficients $a_i$, $i=0,\dots,n$ and the continuous function $a(\tau)$ in the \gls{BC} \eqref{eq:flat_hccf_bc} are characteristic of the system (cf.\ \cite{Russell1991}), similar to the characteristic coefficients of a finite-dimensional \gls{CCF} \eqref{eq:ccf_finite}.
They follow from applying the transformation between the states $\vect\eta(t)$, $\eta_{n+1}(\tau,t)$ and $\vect\xi(t)$, $\bar{\vect x}(z,t)$ to the system representation \eqref{eq:sys_transp}.
This \gls{HCCF} state transformation is detailed in Appendix~\ref{sec:app_hccf_trafo}.
In order to illustrate the underlying idea of the transformation, the following simple example considers a hyperbolic PDE without an ODE subsystem.

\myvspace{.25cm}
\begin{example}[\gls{HCCF} for $n=0$]
\label{rem:flat_HCCF_noODE}
    For a hyperbolic PDE system without an ODE, the \gls{HCCF} \eqref{eq:flat_hccf} only involves the transport PDE
    \begin{subequations}
    \label{eq:flat_hccf_n0}
        \begin{align}
            \partial_t\eta_1(\tau,t) &= \partial_\tau\eta_1(\tau,t), \qquad \tau\in[-\tau_2,\tau_1) \\
        \label{eq:flat_hccf_n0_pde}
            \eta_1(\tau_1,t) &= - \tint_{-\tau_2}^{\tau_1} a(\tau)\eta_1(\tau,t)\,\dd\tau - a_0\eta_1(-\tau_2,t) + \bar u(t)
        \end{align}
    \end{subequations}
    for the state $\eta_1(\tau,t)=y(t+\tau)$, $\tau\in[-\tau_2,\tau_1]$ (cf.\ \eqref{eq:flat_hccf_state} and \cite{Russell1978JMAA}).
    With the boundary value $y(t) = x_1(0,t)$ taking the role of a flat output, the solution \eqref{eq:flat_sol_pde} of the Cauchy problem reads
    \begin{subequations}
    \label{eq:flat_sol_pde_n0}
        \begin{align}
            \bar x_1(z,t) &= \bar x_1(0,t+\phi_1(z)) = \eta_1(\phi_1(z),t) \\
            \bar x_2(z,t) &= \bar x_2(0,t-\phi_2(z)) = q_0 \eta_1(-\phi_2(z),t).
        \end{align}
    \end{subequations}
    As $q_0\neq0$ by Assumption~\ref{enum:ass_pde}, this PDE solution represents the transformation between the \gls{HCCF} state and $\bar{\vect x}(z,t)$, and vice versa.
    For a fixed time $t$, Figure~\ref{fig:HCCFtrafo_n0} illustrate the relation of the spatial domain $[0,1]$ of $\bar{\vect x}(z,t)$ and the time interval $[-\tau_2,\tau_1]$ of $\eta_1(\tau,t)$.
    Based on $\phi_1(z)$ and $\phi_2(z)$ in \eqref{eq:flat_sol_pde_n0}, the graphs (in orange) in Figure~\ref{fig:HCCFtrafo_n0} correspond to characteristic curves of the hyperbolic PDE.
    Moreover, in view of the input parametrization
    \begin{equation}
        \bar u(t) = y(t+\tau_1) - q_0\bar q_1 y(t-\tau_2)
    \end{equation}
    (cf.~\eqref{eq:flat_baru_para}), it is easy to see that $a(\tau)=0$ and $a_0=-q_0\bar q_1$ in the \gls{HCCF} \eqref{eq:flat_hccf_n0}.
    The input parametrization can be interpreted as a neutral delay system, which is stable if $|q_0\bar q_1|<1$ (see \cite{Fridman2014}).
    
    \begin{figure}[t]
    \centering
        \includegraphics{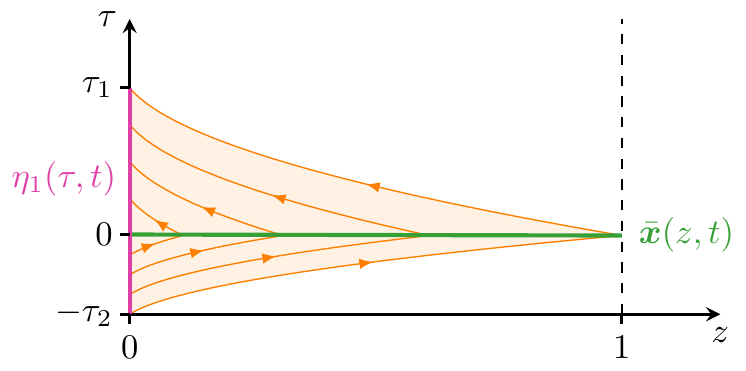}
        \caption{Hyperbolic PDE without an ODE ($n=0$): The relation \eqref{eq:flat_sol_pde_n0} between the \gls{HCCF} state $\eta_1(\tau,t)$ and the PDE state $\bar{\vect x}(z,t)$ of \eqref{eq:sys_transp} for a fixed $t$ is visualized by the graphs in orange that correspond to characteristic curves of the PDE.}
        \label{fig:HCCFtrafo_n0}
    \end{figure}
\end{example}

Alternatively, $a_i$, $i=0,\dots,n$ and $a(\tau)$ can be determined based on the input parametrization, which is simply a representation of the dynamics \eqref{eq:sys_transp} as a time-delay system.
Applying the general relations \eqref{eq:flat_uglyeq1} and \eqref{eq:flat_uglyeq2} from  Appendix~\ref{sec:app_hccf_trafo} to \eqref{eq:flat_baru_para}, the input parametrization reads
\begin{align}
\label{eq:flat_baru_hccf}
    \bar u(t) &= y^{(n)}(t+\tau_1) + a_ny^{(n)}(t-\tau_2) \\
    &\qquad + \sum_{i=0}^{n-1} a_iy^{(i)}(t-\tau_2) + \tint_{-\tau_2}^{\tau_1} a(\tau)y^{(n)}(t+\tau)\,\dd\tau, \nonumber
\end{align}
where all terms on the right-hand side could easily be substituted using the definition \eqref{eq:flat_hccf_state} of the \gls{HCCF} state.
Importantly, \eqref{eq:flat_baru_hccf} can always be solved for $y^{(n)}(t+\tau_1)$ due to the definition of $\vect p_1^\top$ in \eqref{eq:flat_parametrization_z0_x10} and thus directly yields the \gls{BC} \eqref{eq:flat_hccf_bc} in view of $y^{(n)}(t+\tau_1)=\eta_{n+1}(\tau_1,t)$.
While $a_n=-q_0\bar q_1$ is obvious from $\vect p_2^\top$ in \eqref{eq:flat_parametrization_z0_x20}, the explicit definition of $a_i$, $i=0,\dots,n-1$, and $a(\tau)$ is omitted, as the implementation of the flatness-based tracking controller is ultimately not reliant on them.

Let $\vect\eta(0)=\vect\eta_0\in\Rset^n$ and $\eta_{n+1}(\tau,0)=\eta_{n+1,0}(\tau)\in\Rset$ be the \glspl{IC} of \eqref{eq:flat_hccf} implied by the \glspl{IC} of \eqref{eq:sys_transp} and the \gls{HCCF} state transformation (see Lemma~\ref{lem:hccf_trafo} in Appendix~\ref{sec:app_hccf_trafo}).
Then, the following result is consequence of Lemmas~\ref{lem:equiv_sys_pre} and \ref{lem:hccf_trafo}.

\begin{lem}[\gls{HCCF}]
\label{lem:hccf}
    The \gls{HCCF} \eqref{eq:flat_hccf} with the state \eqref{eq:flat_hccf_state} is an equivalent representation of every system \eqref{eq:sys} that satisfies Assumptions \ref{enum:ass_ode} and \ref{enum:ass_pde}.
\end{lem}

It is noteworthy that the stability of the PDE-ODE system with its equivalent representations \eqref{eq:sys_transp}, \eqref{eq:flat_hccf} and \eqref{eq:flat_baru_hccf} is implied by the solutions of the characteristic equation
\begin{equation}
\label{eq:flat_sys_chareq}
    0 = \mathfrak a(s) = s^n\!\!\left[\rme^{\tau_1s} + a_n\rme^{-\tau_2s} + \!\!\tint_{-\tau_2}^{\tau_1} a(\tau)\rme^{\tau s}\,\dd\tau\right] + \sum_{i=0}^{n-1} a_i s^i
\end{equation}
(see, e.g., \cite{Fridman2014} or \cite[Thm.~4.1]{Russell1991}), where $\mathfrak a$ is apparent from both \eqref{eq:flat_hccf} and \eqref{eq:flat_baru_hccf}.
It is shown in \cite[Thm.~4.1]{Russell1991} that \eqref{eq:flat_hccf} is a Riesz spectral system if $a_n\neq0$, assuming an appropriate state space for \eqref{eq:flat_hccf_state}.
While the location of the roots $s_i$, $i\in\Nset$, of $\mathfrak a$ is insufficient to deduce stability in general, for Riesz spectral systems, $\sup_{i\in\Nset}\re(s_i)<0$ ensures exponential stability (see \cite[Thm.~3.2.8]{Curtain2020}).
For the stabilization of \eqref{eq:flat_hccf}, the characteristic function $\mathfrak a$ would basically be replaced by a desired one for the closed loop, 
analogous to the finite-dimensional setting in \eqref{eq:ccf_finite}.
However, while this is the underlying idea of the following control design, Section~\ref{sec:flatness_tracking} will also make it clear that the \gls{HCCF} is not explicitly required.

\subsection{Tracking controller}
\label{sec:flatness_tracking}

In order to design a tracking controller, first, an appropriate reference is specified.

\subsubsection{Reference trajectories and feedforward controller}
\label{sec:reference}

It is shown in Section~\ref{sec:flat_para} that the flat output \eqref{eq:flat_output} parametrizes all system variables of \eqref{eq:sys_transp} (cf.\ \eqref{eq:flat_parametrization_z0_xi}, \eqref{eq:flat_x_para} and \eqref{eq:flat_baru_para}).
Therefore, instead of specifying a reference for $\vect\xi(t)$ and $\bar{\vect x}(z,t)$ that satisfies \eqref{eq:sys_transp}, an arbitrary (but sufficiently smooth) desired trajectory for $y(t)$ is chosen.
For instance, the transition of a system between two steady states in arbitrary, finite time $t_*-t_0$ can be characterized by a piecewise-defined reference function
\begin{equation}
\label{eq:ysoll}
    y_\soll(t) = \begin{cases}
        y_0, & t < t_0 \\
        \Xi(t), & t \in [t_0,t_*] \\
        y_*, & t > t_*.
    \end{cases}
\end{equation}
As all time derivatives of $y_\soll(t)$ vanish for $t<t_0$ and $t>t_*$, the chosen initial and terminal values, $y_0$ and $y_*$, imply stationary solutions for $\vect\xi(t)$ and $\bar{\vect x}(z,t)$ by the flatness-based parametrizations \eqref{eq:flat_parametrization_z0_xi}, \eqref{eq:flat_x_para} and \eqref{eq:flat_baru_para}.
Conversely, an equilibrium solution of \eqref{eq:sys_transp} can be reformulated in terms of $y(t)$ using the \gls{HCCF} state transformation (see Appendix~\ref{sec:app_hccf_trafo} and Lemma~\ref{lem:hccf_trafo}).
In view of the $n$-dimensional ODE state $\vect\xi(t)$, the function $\Xi(t)$ is oftentimes chosen as a polynomial
\begin{equation}
\label{eq:ysoll_polynomial}
    \Xi(t) = \sum_{i=0}^{2n+1} c_i \left(\frac{t-t_0}{t_*-t_0}\right)^i
\end{equation}
of degree at least $2n+1$ so that $y_\soll^{(n)}(t)$ is continuous for all $t$ (e.g.~\cite{Levine2009,Rudolph2021}).
The coefficients $c_i$, $i=0,\dots,2n+1$, are computed by requiring
\begin{subequations}
    \begin{align}
        y_\soll(t_0) &= y_0, & y_\soll^{(i)}(t_0) &= 0, & i &= 1,\dots,n \\
        y_\soll(t_*) &= y_*, & y_\soll^{(i)}(t_*) &= 0, & i &= 1,\dots,n.
    \end{align}
\end{subequations}
They are independent of the times $t_0$ and $t_*$ as a result of the special polynomial structure in \eqref{eq:ysoll_polynomial}.
In the end, by the choice of a polynomial \eqref{eq:ysoll_polynomial} of degree $2n+1$, the transition time $t_*-t_0$, the initial value $y_0$ and the terminal value $y_*$ remain as design parameters for the reference trajectory.

Based on $y_\soll(t)$, references for all system variables in \eqref{eq:sys_transp} are found by substituting $y_\soll(t)$ for $y(t)$ in their respective flatness-based parametrizations.
This, in turn, yields the references \eqref{eq:sys_reference} by Lemma~\ref{lem:equiv_sys_pre}.

\myvspace{.25cm}
\begin{lem}[References]
\label{lem:references}
    The references $\vect\xi_\soll(t)$, $\bar{\vect x}(z,t)$, $\bar u_\soll(t)$ implied by $y_\soll(t)$ and \eqref{eq:flat_parametrization_z0_xi}, \eqref{eq:flat_x_para}, \eqref{eq:flat_baru_para} satisfy \eqref{eq:sys_transp}.
\end{lem}

\begin{figure}
    \centering
    \includegraphics{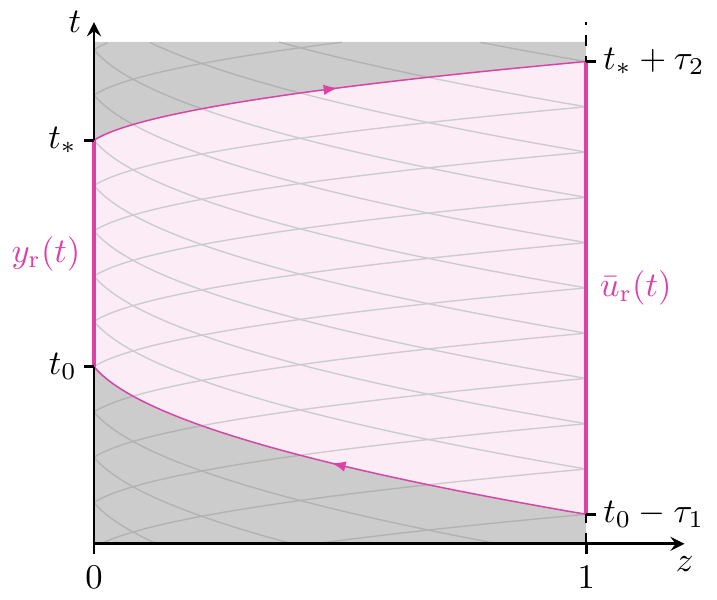}
    \caption{Characteristic curves of the PDE subsystem illustrate the relation between the transition times for $y_\soll(t)$ and $\bar u(t)$. Gray areas correspond to steady states.}
    \label{fig:reference}
\end{figure}

The result is immediately proven since \eqref{eq:flat_parametrization_z0_xi}, \eqref{eq:flat_x_para} and \eqref{eq:flat_baru_para} parameterize all system solutions of \eqref{eq:sys_transp} (see \cite{Woittennek2013CPDE}).
Substituting $y(t+\tau)=y_\soll(t+\tau)$ for $\tau\in[-\tau_2,\tau_1]$ in the parametrization \eqref{eq:flat_baru_para} or equivalently \eqref{eq:flat_baru_hccf} yields the reference
\begin{align}
\label{eq:flat_feedforward}
    \bar u_\soll(t) &= y_\soll^{(n)}(t+\tau_1) + a_ny_\soll^{(n)}(t-\tau_2) \\
    &\qquad + \sum_{i=0}^{n-1} a_iy_\soll^{(i)}(t-\tau_2) + \tint_{-\tau_2}^{\tau_1} a(\tau)y_\soll^{(n)}(t+\tau)\,\dd\tau \nonumber
\end{align}
for the input $\bar u(t)$ and thus a feedforward controller for \eqref{eq:sys_transp}.
Due to the time shift in \eqref{eq:flat_baru_para}, a transition time $t_*-t_0$ for $y_\soll(t)$ between two steady states results in a non-constant $\bar u_\soll(t)$ for $t\in[t_0-\tau_1,t_*+\tau_2]$.
This is illustrated by the characteristic curves in Figure~\ref{fig:reference} that showcase that a control action at $z=1$ takes the time $\tau_1$ to affect the opposite boundary at $z=0$ where the reference for the flat output is specified.
Thus, control action is necessary before the start of the transition at $t_0$.
In turn, post action on $[t_*,t_*+\tau_2]$ in $\bar u_\soll(t)$ is required to account for the transport delay $\tau_2$ in the positive $z$-direction.
Ultimately, while the set-point change for the ODE subsystem and all quantities at $z=0$ takes place in $[t_0,t_*]$, the overall system is in steady state only for $t<t_0-\tau_1$ and $t>t_*+\tau_2$.

References $\vect\xi_\soll(t)$ and $\bar{\vect x}_\soll(z,t)$ directly follow from replacing $y(t)$ in \eqref{eq:flat_parametrization_z0_xi} and \eqref{eq:flat_x_para} by $y_\soll(t)$.
They in turn give the remaining references in \eqref{eq:sys_reference}.

\subsubsection{Controller and closed loop}

\begin{figure*}[t]
    \centering
    \scalebox{.97}{\includegraphics{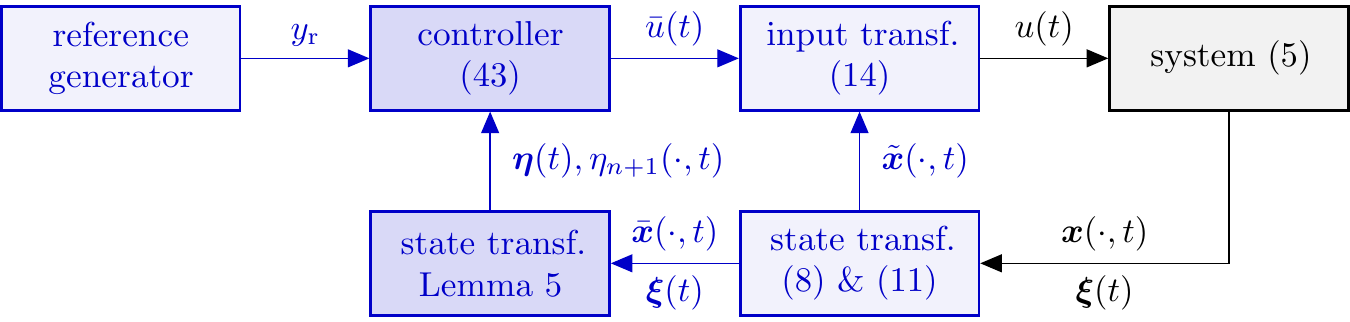}}
    \caption{Signal flow diagram of the closed-loop system using the flatness-based controller \eqref{eq:flat_statefeedback} to track a reference \eqref{eq:sys_reference} implied by $y_\soll(t)$ for the PDE-ODE system \eqref{eq:sys}.
    Blocks with a stronger saturation highlight the main design components and the differences to the backstepping-based controller in Figure~\ref{fig:bs-signalflow}.}
    \label{fig:flat-signalflow}
\end{figure*}

Analogous to the control design based on the \gls{CCF} \eqref{eq:ccf_finite} for finite-dimensional systems, a stabilizing controller for the \gls{HCCF} \eqref{eq:flat_hccf} replaces the coefficients $a_i$, $i=0,\dots,n$, and the function $a(\tau)$ by desired ones that are associated with a stable closed-loop dynamics.
Augmenting such a feedback with the feedforward part $\bar u_\soll(t)$ in \eqref{eq:flat_feedforward} allows to track the reference $y_\soll(t)$.

In view of the relation between the characteristic equation \eqref{eq:flat_sys_chareq} of the system and the input parametrization \eqref{eq:flat_baru_hccf} as well as the \gls{BC} \eqref{eq:flat_hccf_bc} of the \gls{HCCF}, the choice of a desired closed-loop dynamics translates into specifying an appropriate functional differential equation for the tracking error $e_y(t)=y(t)-y_\soll(t)$.
With the general case discussed in Remark~\ref{rem:flat_cl}, for simplicity, the infinite-dimensional error dynamics
\begin{equation}
\label{eq:flat_closedloop}
    0 = \epsilon^{(n)}(t) + \sum_{i=0}^{n-1} \kappa_i \epsilon^{(i)}(t)
\end{equation}
is considered where $\epsilon(t)=e_y(t+\tau_1)+\gamma e_y(t-\tau_2)$.
From the corresponding characteristic equation
\begin{equation}
\label{eq:flat_sys_chareq_closedloop}
    0 = \mathfrak a_{\mathrm{cl}}(s) = (\rme^{\tau_1s} + \gamma\rme^{-\tau_2s}) \left[ s^n + \sum_{i=0}^{n-1} \kappa_i s^i\right],
\end{equation}
it can easily be verified by explicit calculation that the choice $0<|\gamma|<1$ with coefficients $\kappa_i$, $i=0,\dots,n-1$, of a Hurwitz polynomial results in $\re(s_i)<0$, $i\in\Nset$, for the roots of $\mathfrak a_{\mathrm{cl}}$.
Recalling that the \gls{HCCF} \eqref{eq:flat_hccf} is a Riesz spectral system (see text below \eqref{eq:flat_sys_chareq} and \cite[Thm.~4.1]{Russell1991}), these parameters ensure an exponentially stable closed-loop system.
Moreover, exponential stability can even be shown for $|\gamma|<1$ and coefficients $\kappa_i$, $i=0,\dots,n-1$, of a Hurwitz polynomial, as the solution of \eqref{eq:flat_closedloop} decays exponentially for $t>\tau_1$ if $\gamma=0$.

The link between the open-loop system represented by \eqref{eq:flat_baru_hccf} and the closed loop \eqref{eq:flat_closedloop} becomes clearer if \eqref{eq:flat_uglyeq2} in Appendix \ref{sec:app_hccf_trafo} is adapted to $e_y^{(i)}(t+\tau_2)$, $i=0,\dots,n-1$, in order to rewrite \eqref{eq:flat_closedloop} in a form
\begin{align}
\label{eq:flat_closedloop_rewrite}
    0 &= e_y^{(n)}(t+\tau_1) + \bar a_n e_y^{(n)}(t-\tau_2) \\
    &\qquad + \sum_{i=0}^{n-1} \bar a_i e_y^{(i)}(t-\tau_2) + \int_{-\tau_2}^{\tau_1} \bar a(\tau) e_y^{(n)}(t+\tau)\,\dd\tau \nonumber
\end{align}
analogous to \eqref{eq:flat_baru_hccf}, with
\begin{subequations}
    \begin{align}
        \bar a(\tau) &= \sum_{i=0}^{n-1} \kappa_i \frac{(\tau_1-\tau)^{n-i-1}}{(n-i-1)!} \\
        \bar a_i &= \kappa_i\gamma + \sum_{k=0}^{i} \kappa_k \frac{(\tau_1+\tau_2)^{i-k}}{(i-k)!}
    \end{align}
\end{subequations}
for $i=0,\dots,n-1$ and $\bar a_n=\gamma$.
While the representation \eqref{eq:flat_closedloop} of the closed loop is advantageous for the stability analysis, it is more apparent from \eqref{eq:flat_closedloop_rewrite} that the open-loop parameters $a_i$, $i=0,\dots,n$, and $a(\tau)$ will be replaced by some desired ones $\bar a_i$, $i=0,\dots,n$ and $\bar a(\tau)$ in the controlled system.
This is verified based on the flatness-based tracking controller
\begin{align}
\label{eq:flat_controller}
    \bar u(t) &= \bar u_\soll(t) + (a_n-\bar a_n) e_y^{(n)}(t-\tau_2) \nonumber \\
    &\qquad + \sum_{i=0}^{n-1} (a_i-\bar a_i) e_y^{(i)}(t-\tau_2) \nonumber \\
    &\qquad + \int_{-\tau_2}^{\tau_1} \big(a(\tau)-\bar a(\tau)\big) e_y^{(n)}(t+\tau)\,\dd\tau 
\end{align}
that follows from solving \eqref{eq:flat_closedloop_rewrite} for $y^{(n)}(t+\tau_1)$, i.e.\ the boundary value of the \gls{HCCF} \eqref{eq:flat_hccf}, and substituting the result into \eqref{eq:flat_baru_hccf} together with \eqref{eq:flat_feedforward}.
Hence, applying \eqref{eq:flat_controller} to \eqref{eq:flat_baru_hccf} yields the desired closed loop \eqref{eq:flat_closedloop_rewrite}.
Even though the integral in \eqref{eq:flat_controller} involves positive time shifts, with the reference $y_\soll(t)$ known in advance, no predictions are required for the implementation of \eqref{eq:flat_controller}.
In fact, in view of the definition of the tracking error $e_y(t)$ and the \gls{HCCF} state \eqref{eq:flat_hccf_state}, \eqref{eq:flat_controller} is easily be rewritten as the state feedback
\begin{align}
\label{eq:flat_statefeedback}
    \bar u(t) &= \bar u_\soll(t) + (a_n-\bar a_n) \big(\eta_{n+1}(-\tau_2,t)-y_\soll^{(n)}(t-\tau_2) \big) \nonumber \\
    &\qquad + \sum_{i=0}^{n-1} (a_i-\bar a_i) \big(\eta_{i+1}(t)-y_\soll^{(i)}(t-\tau_2)\big) \\
    &\qquad + \int_{-\tau_2}^{\tau_1} \big(a(\tau)-\bar a(\tau)\big) \big(\eta_{n+1}(\tau,t)-y_\soll^{(n)}(t+\tau)\big)\,\dd\tau \nonumber
\end{align}
based on $\vect\eta(t)$ and $\eta_{n+1}(\tau,t)$.
The overall closed-loop system is visualized by the signal flow diagram in Figure~\ref{fig:flat-signalflow}.

The following theorem asserts the stability of the closed-loop dynamics achieved by the flatness-based controller \eqref{eq:flat_statefeedback} and thus the tracking of a desired reference trajectory.

\begin{thm}[Closed-loop stability]
\label{thm:flatness}
    Let Assumptions \ref{enum:ass_ode} and \ref{enum:ass_pde} hold and $|\gamma|<1$ with coefficients $\kappa_i$, $i=0,\dots,n-1$, of a Hurwitz polynomial.
    Then, the state feedback controller \eqref{eq:flat_statefeedback} exponentially stabilizes the system \eqref{eq:sys} along reference trajectories \eqref{eq:sys_reference}.
\end{thm}

The result directly follows from Lemmas~\ref{lem:equiv_sys_pre}--\ref{lem:references}.
Based on Assumptions \ref{enum:ass_ode} and \ref{enum:ass_pde}, the \gls{HCCF} \eqref{eq:flat_hccf} is an equivalent state-space representation of \eqref{eq:sys} by Lemma \ref{lem:equiv_sys_pre} and Lemma \ref{lem:hccf}.
The feedback \eqref{eq:flat_statefeedback} of the \gls{HCCF} state ensures convergence of the flat output $y(t)$ towards its reference $y_\soll(t)$ (cf.\ \eqref{eq:flat_closedloop}), which in turn implies the same for the references \eqref{eq:sys_reference} by Lemma \ref{lem:references} and Lemma \ref{lem:equiv_sys_pre}, thus completing the proof.
Theorem~\ref{thm:flatness} holds analogously for any other appropriate choice of closed-loop dynamics (see Remark \ref{rem:flat_cl}).
The closed loop \eqref{eq:flat_closedloop} is chosen for simplicity of presentation only.

\begin{rem}[Choice of closed-loop dynamics]
\label{rem:flat_cl}
    The flatness-based design allows for a very general choice of closed-loop dynamics.
    For example, any functional differential equation
    \begin{multline}
    \label{eq:flat_closedloop_allg}
        0 = \sum_{i=0}^n \left(\tilde a_i^+ e_y^{(i)}(t+\tau_1) + \tilde a_i^- e_y^{(i)}(t-\tau_2)\right. \\
        \left.+ \int_{-\tau_2}^{\tau_1} \tilde a_i(\tau)e_y^{(i)}(t+\tau)\,\dd\tau\right), \qquad \tilde a_n^+ = 1,
    \end{multline}
    with design parameters $\tilde a_i^+$, $\tilde a_i^-$, $\tilde a_i(\tau)$, $i=0,\dots,n$ such that the tracking error $e_y(t)$ converges to zero can be chosen.
    Importantly, $\tilde a_n^+=1$ ensures that \eqref{eq:flat_closedloop_allg} can be solved for $y^{(n)}(t+\tau_1)$ to obtain the controller corresponding to \eqref{eq:flat_closedloop_allg} by substitution into \eqref{eq:flat_baru_hccf}.
    Note that \eqref{eq:flat_closedloop_allg} has to be at least of order $n$, with a choice of order larger than $n$ resulting in a dynamic feedback, as opposed to the static one in \eqref{eq:flat_statefeedback}.
    Moreover, it is even possible to choose a nonlinear functional differential equation for the closed loop.
\end{rem}

The presented derivation of the state feedback \eqref{eq:flat_statefeedback} emphasizes the parallels between the controller design based on the finite-dimensional \gls{CCF} \eqref{eq:ccf_finite} and the \gls{HCCF} \eqref{eq:flat_hccf}, respectively.
However, it is possible to do without the \gls{HCCF} and thus without explicitly determining $a_i$, $i=0,\dots,n$, and $a(\tau)$ in \eqref{eq:flat_hccf}, which may be advantageous for the implementation of a flatness-based tracking controller.
Those familiar with the flatness-based control design for finite-dimensional systems already know that a stabilizing feedback is usually directly determined based on an input parametrization  (e.g.~\cite{Levine2009,Rudolph2021}).
This too is possible for the \glspl{DPS} considered here, where both the \gls{HCCF} and the controller directly follow on the basis of the input parametrization.
Looking at \eqref{eq:flat_baru_para}, it is interesting to observe that the restriction of the trajectory of the flat output $y(t)$ to the interval $[t-\tau_2,t+\tau_1]$ constitutes a state of the system at time $t$ in an appropriate state space\footnote{The restriction belongs to the Sobolev space $H^n([-\tau_2,\tau_1])$ of $n$-times weakly differentiable functions in $L^2([-\tau_2,\tau_1])$, which is isomorphic to the \gls{HCCF} state space $\Rset^n\times L^2([-\tau_2,\tau_1])$ (see, e.g., \cite[Sec. 5.1]{Woittennek2013CPDE} for details).}.
As such, solving \eqref{eq:flat_closedloop} for $y^{(n)}(t+\tau_1)$ and substituting the result in the input parametrization \eqref{eq:flat_baru_para} already yields a state feedback (see \cite{Woittennek2013CPDE}).
Using the transformation in Appendix \ref{sec:app_hccf_trafo} (see also Lemma~\ref{lem:hccf_trafo}) as well as the preliminary transformations \eqref{eq:pre_trafo_hopfcole} and \eqref{eq:pre_trafo_bs}, it can be written as a feedback of the states $\vect x(z,t)$ and $\vect\xi(t)$ of \eqref{eq:sys}.
It should be noted that this abbreviated control design better reflects the solution-oriented perspective of the flatness-based approach, which has a strong connection to the behavioral approach in, e.g., \cite{Willems1991tac}.
However, in contrast to the algorithmic design based on the \gls{HCCF}, such an approach requires a rigorous discussion of state spaces, which is largely omitted in this paper to keep the presentation as simple as possible.


  \section{Backstepping-based design}
\label{sec:backstepping}

In order to design a tracking controller for \eqref{eq:sys_transp} using backstepping, the control problem is reformulated as a stabilization task.
For that, introduce the tracking errors
\begin{equation}
\label{eq:bs_error_def}
    \vect e_\xi(t) = \vect\xi(t)-\vect\xi_\soll(t) \quad\text{and}\quad \vect\varepsilon(z,t) = \bar{\vect x}(z,t)-\bar{\vect x}_\soll(z,t)
\end{equation}
that satisfy the dynamics
\begin{subequations}
\label{eq:bs_error_sys}
    \begin{align}
    \label{eq:bs_error_sys_ode}
        \dot{\vect e}_\xi(t) &= \mat F\vect e_\xi(t) + \vect b \varepsilon_1(0,t) \\
    \label{eq:bs_error_sys_bc0}
        \varepsilon_2(0,t) &= q_0 \varepsilon_1(0,t) + \vect c^\top \vect e_\xi(t) \\
    \label{eq:bs_error_sys_pde}
        \partial_t \vect\varepsilon(z,t) &= \mat\Lambda(z) \partial_z \vect\varepsilon(z,t) + \mat C(z)\vect e_\xi(t) \\
    \label{eq:bs_error_sys_bc1}
        \varepsilon_1(1,t) &= \bar q_1\varepsilon_2(1,t) + \bar u(t) - \bar u_\soll(t)
    \end{align}
\end{subequations}
based on \eqref{eq:sys_transp} as well as references $\vect\xi_\soll(t)$, $\bar{\vect x}_\soll(z,t)$ and $\bar u_\soll(t)$ that solve \eqref{eq:sys_transp} (e.g.\ flatness-based references, see Lemma~\ref{lem:references}).
The \glspl{IC} $\vect e_\xi(0)=\vect e_{\xi,0}= \vect\xi_0-\vect\xi_\soll(0)$, $\vect\varepsilon(z,0)=\vect\varepsilon_0(z)=\bar{\vect x}_0(z)-\bar{\vect x}_\soll(z,0)$ are defined accordingly.

\begin{figure}
    \centering
    \includegraphics[width=\columnwidth]{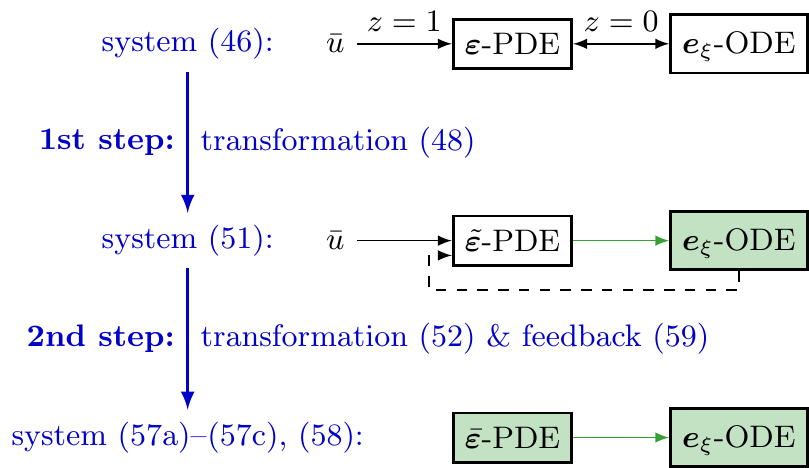}
    \caption{Overview of the two steps involved in the backstepping design. Green blocks symbolize stabilized subsystems.}
    \label{fig:bs-overview}
\end{figure}

Systems \eqref{eq:sys_transp} and \eqref{eq:bs_error_sys} share the same coupling structure that is illustrated in Figure~\ref{fig:systransp_structure}.
Importantly, both are in strict-feedback form (see \cite{Gehring2021MTNS}) as the ODE subsystem \eqref{eq:bs_error_sys_ode} is driven only by the boundary value $\varepsilon_1(0,t)$ of the PDE, with the control input $\bar u(t)$ acting on the opposite boundary at $z=1$ of the PDE subsystem \eqref{eq:bs_error_sys_bc0}--\eqref{eq:bs_error_sys_bc1}.
This structure is exploited for a recursive control design.
In a first step, the ODE subsystem \eqref{eq:bs_error_sys_ode} is stabilized by an appropriate choice of its virtual input $\varepsilon_1(0,t)$.
Analogous to the backstepping design for finite-dimensional systems, this virtual feedback induces a state transformation into error variables that are to be driven to zero by means of the input $\bar u(t)$ in the second and final step.
Figure~\ref{fig:bs-overview} roughly illustrates the coupling structure between the ODE and PDE subsystem following each of the two design steps.
It is important to note that the method chosen for the stabilization of any of the subsystems is arbitrary.
As such, in the classical sense of \eqref{eq:strictfeedback} where $\eta_1(t)$ relates to $\vect e_\xi(t)$ in \eqref{eq:bs_error_sys} and $\eta_2(t)$ to $\vect\varepsilon(z,t)$, this design strategy involves only the one backstepping step in Section~\ref{sec:bs_ode}.

\subsection{Stabilization of the ODE subsystem}
\label{sec:bs_ode}

First, only the ODE subsystem \eqref{eq:bs_error_sys_ode} is considered, wherein the boundary term $\varepsilon_1(0,t)$ takes the role of an input.
In light of the controllability of the pair $(\mat F,\vect b)$ (see \ref{enum:ass_ode}), it is always possible to choose a (virtual) feedback
\begin{equation}
\label{eq:bs_virtual_feedback}
    \varepsilon_1(0,t) = \vect k^\top\vect e_\xi(t)
\end{equation}
such that $\mat F+\vect b\vect k^\top$ is Hurwitz, i.e., all eigenvalues of the matrix have a negative real part, which ensures exponential stability of $\vect e_\xi(t)$.
However, such a control law cannot be implemented, as $\varepsilon_1(0,t)$ is not an input.
In fact, \eqref{eq:bs_virtual_feedback} may be interpreted as a desired value in the sense that $\varepsilon_1(0,t)\to\vect k^\top\vect e_\xi(t)$ for $t\to\infty$ is to be achieved.

This motivates the definition of the PDE error state
\begin{equation}
\label{eq:bs_trafo_decoupl}
    \tilde{\vect\varepsilon}(z,t) = \vect\varepsilon(z,t) - \mat N(z)\vect e_\xi(t)
\end{equation}
with a matrix $\mat N(z)\in\Rset^{2\times n}$ yet to be determined that satisfies at least $\vect e_1^\top \mat N(0)=\vect k^\top$ in order for the error $\tilde\varepsilon_1(0,t)=0$ to imply \eqref{eq:bs_virtual_feedback}.
Using this condition on $\mat N(z)$, the backstepping transformation \eqref{eq:bs_trafo_decoupl} maps the dynamics \eqref{eq:bs_error_sys} into the form
\begin{subequations}
\label{eq:bs_sys_decoupl_pre}
    \begin{align}
    \label{eq:bs_sys_decoupl_pre_ode}
        \dot{\vect e}_\xi(t) &= (\mat F+\vect b\vect k^\top)\vect e_\xi(t) + \vect b \tilde\varepsilon_1(0,t) \\
    \label{eq:bs_sys_decoupl_pre_bc0}
        \tilde\varepsilon_2(0,t) &= q_0 \tilde\varepsilon_1(0,t) + \left[\vect c^\top+q_0\vect k^\top-\vect e_2^\top\mat N(0)\right]\vect e_\xi(t) \\
    \label{eq:bs_sys_decoupl_pre_pde}
        \partial_t \tilde{\vect\varepsilon}(z,t) &= \mat\Lambda(z) \partial_z \tilde{\vect\varepsilon}(z,t) - \mat N(z)\vect b\tilde\varepsilon_1(0,t) \\
        &\hspace{.15cm} + \left[\mat\Lambda(z)\mat N^\prime(z)-\mat N(z)(\mat F+\vect b\vect k^\top)+\mat C(z)\right]\vect e_\xi(t) \nonumber \\
    \label{eq:bs_sys_decoupl_pre_bc1}
        \tilde\varepsilon_1(1,t) &= \bar q_1\varepsilon_2(1,t) - \vect e_1^\top \mat N(1)\vect e_\xi(t) + \bar u(t) - \bar u_\soll(t),
    \end{align}
\end{subequations}
with the sought  matrix $\mat F+\vect b\vect k^\top$ in \eqref{eq:bs_sys_decoupl_pre_ode}.
The remaining freedom in the choice of $\mat N(z)$ is used to decouple the PDE from the ODE subsystem.
This allows to consider the stabilization of the PDE subsystem \eqref{eq:bs_sys_decoupl_pre_bc0}--\eqref{eq:bs_sys_decoupl_pre_bc1} independent of the ODE subsystem \eqref{eq:bs_sys_decoupl_pre_ode} in Section~\ref{sec:bs_pde}.
By requiring $\mat N(z)$ to satisfy the initial value problem
\begin{subequations}
\label{eq:bs_decouple_ivp}
    \begin{align}
    \label{eq:bs_decouple_ode}
        \mat\Lambda(z)\mat N^\prime(z) &= \mat N(z)(\mat F+\vect b\vect k^\top) - \mat C(z) \\
    \label{eq:bs_decouple_ic}
        \mat N(0) &= \begin{bmatrix} \vect k^\top \\ \vect c^\top + q_0\vect k^\top \end{bmatrix}
    \end{align}
\end{subequations}
with \eqref{eq:bs_decouple_ode} for $z\in(0,1]$, the ODE state $\vect e_\xi(t)$ only impacts the actuated boundary \eqref{eq:bs_sys_decoupl_pre_bc1}.
There, it can (and will) be compensated by an appropriate choice of the control input $\bar u(t)$.
For the same reason, the right-hand side of the actuated boundary \eqref{eq:bs_sys_decoupl_pre_bc1} is not transformed into the new coordinates.
Motivated by its purpose to decouple the PDE from the ODE subsystem, \eqref{eq:bs_trafo_decoupl} is also referred to as a decoupling transformation, \eqref{eq:bs_decouple_ivp} as the decoupling equations (e.g.\ \cite{Deutscher2019ijc}).

The solution of \eqref{eq:bs_decouple_ivp} is found analogously to \cite[Thm.~4.1]{Deutscher2019ijc} by multiplying \eqref{eq:bs_decouple_ivp} with the eigenvectors of $\mat F+\vect b\vect k^\top$ from the right.
This breaks down the matrix-valued initial value problem into $n$ vector-valued ones with varying coefficients due to $\mat\Lambda(z)$, which can be solved explicitly\footnote{This is due to the preliminary transformations in Section~\ref{sec:prelim}, as applying the decoupling transformation to \eqref{eq:sys} directly would result in an initial value problem that cannot be solved explicitly.}.
These unique, continuously differentiable solutions (e.g.\ \cite[Thm.~3.3]{Rugh1996}) verify the existence of a unique solution $\mat N\in (C^1([0,1]))^{2\times n}$ for \eqref{eq:bs_decouple_ivp}.
Consequently, the backstepping transformation \eqref{eq:bs_trafo_decoupl} maps \eqref{eq:bs_error_sys} into 
\begin{subequations}
\label{eq:bs_sys_decoupl}
    \begin{align}
    \label{eq:bs_sys_decoupl_ode}
        \dot{\vect e}_\xi(t) &= (\mat F+\vect b\vect k^\top)\vect e_\xi(t) + \vect b \tilde\varepsilon_1(0,t) \\
    \label{eq:bs_sys_decoupl_bc0}
        \tilde\varepsilon_2(0,t) &= q_0 \tilde\varepsilon_1(0,t) \\
    \label{eq:bs_sys_decoupl_pde}
        \partial_t \tilde{\vect\varepsilon}(z,t) &= \mat\Lambda(z) \partial_z \tilde{\vect\varepsilon}(z,t) - \mat N(z)\vect b\tilde\varepsilon_1(0,t) \\
    \label{eq:bs_sys_decoupl_bc1}
        \tilde\varepsilon_1(1,t) &= \bar q_1\varepsilon_2(1,t) - \vect e_1^\top \mat N(1)\vect e_\xi(t) + \bar u(t) - \bar u_\soll(t),
    \end{align}
\end{subequations}
which is simply \eqref{eq:bs_sys_decoupl_pre} in light of the conditions \eqref{eq:bs_decouple_ivp} on $\mat N(z)$.
Importantly, the transformation \eqref{eq:bs_trafo_decoupl} corresponds to a classical backstepping step in the sense of \eqref{eq:strictfeedback} and preserves the strict-feedback form in \eqref{eq:bs_sys_decoupl}, with stability of the ODE subsystem \eqref{eq:bs_sys_decoupl_ode} ensured by the choice of $\vect k$.
As \eqref{eq:bs_sys_decoupl} only involves a cascade of a PDE subsystem and a stable ODE subsystem (see Figure~\ref{fig:bs-overview}), at least if $\bar u(t)$ compensates the ODE state in \eqref{eq:bs_sys_decoupl_bc1}, stabilizing the PDE subsystem in the next step guarantees an overall stable system.
Note that the method chosen for the stabilization of \eqref{eq:bs_sys_decoupl_bc0}--\eqref{eq:bs_sys_decoupl_bc1} is arbitrary.

\myvspace{.75cm}
\subsection{Stabilization of the PDE subsystem}
\label{sec:bs_pde}

The previous transformation \eqref{eq:bs_trafo_decoupl} introduced the local term $-\mat N(z)\vect b\tilde\varepsilon_1(0,t)$ in \eqref{eq:bs_sys_decoupl_pde} that may have a destabilizing effect on the PDE subsystem.
As a consequence, the choice of a stabilizing feedback is not obvious from \eqref{eq:bs_sys_decoupl_bc0}--\eqref{eq:bs_sys_decoupl_bc1}.
In order to facilitate the stabilization of the PDE subsystem, the Volterra integral transformation
\begin{equation}
\label{eq:bs_trafo_bs}
    \bar{\vect\varepsilon}(z,t) = \tilde{\vect\varepsilon}(z,t) - \tint_0^z \mat P(z,\zeta)\tilde{\vect\varepsilon}(\zeta,t)\,\dd\zeta
\end{equation}
with $\mat P(z,\zeta)\in\Rset^{2\times2}$ on the triangular domain $\mathcal T$ in \eqref{eq:triangulardomein} is used to recover the simple transport equations
\begin{equation}
\label{eq:bs_pre_transportpde}
    \partial_t \bar{\vect\varepsilon}(z,t) = \mat\Lambda(z) \partial_z \bar{\vect\varepsilon}(z,t)
\end{equation}
in the new coordinates.

For that, analogous to the preliminary transformation \eqref{eq:pre_trafo_bs} and the calculations done in the context of \eqref{eq:pre_pde_bilanz}, \eqref{eq:bs_trafo_bs} is differentiated w.r.t.\ $t$ and $z$.
Substituting $\partial_t \tilde{\vect\varepsilon}(z,t)$ by means of \eqref{eq:bs_sys_decoupl_pde} and using an integration by parts together with the \gls{BC} \eqref{eq:bs_sys_decoupl_bc0}, it is revealed that \eqref{eq:bs_trafo_bs} maps \eqref{eq:bs_sys_decoupl_pde} into \eqref{eq:bs_pre_transportpde} if the kernel $\mat P(z,\zeta)$ satisfies
\begin{subequations}
\label{eq:bs_kerneleqs}
    \begin{align}
    \label{eq:bs_kerneleqs_pde}
        \mat\Lambda(z)\partial_z \mat P(z,\zeta) + \partial_\zeta\big(\mat P(z,\zeta)\mat\Lambda(\zeta)\big) &= \mat 0 \\
    \label{eq:bs_kerneleqs_bcz}
        \mat P(z,z)\mat\Lambda(z) - \mat\Lambda(z)\mat P(z,z) &= \mat 0 \\
    \label{eq:bs_kerneleqs_bc0}
        \mat P(z,0)\mat\Lambda(0)(\vect e_1+q_0\vect e_2) + \!\tint_0^z \mat P(z,\zeta)\mat N(\zeta)\vect b\,\dd\zeta &= \mat N(z)\vect b,
    \end{align}
\end{subequations}
with \eqref{eq:bs_kerneleqs_pde} defined on the triangular domain $\mathcal T$.
In contrast to \eqref{eq:pre_kerneleqs}, the kernel equations \eqref{eq:bs_kerneleqs} involve an integral \gls{BC} for $\mat P(z,\zeta)$ at $\zeta=0$.
By tracing \eqref{eq:bs_kerneleqs} back to Volterra integral equations, it is shown in \cite[Lem.~6]{Deutscher2018AUT} that this type of kernel equations admits a unique piecewise continuously differentiable solution.
In fact, the solution is straightforward for the simple case of a $2\times2$ matrix $\mat P(z,\zeta)$ considered here. 
Based on the four scalar transport equations in \eqref{eq:bs_kerneleqs_pde}, the two \glspl{BC} $P_{12}(z,z)=P_{21}(z,z)=0$ contained in \eqref{eq:bs_kerneleqs_bcz} yield $P_{12}(z,\zeta)=P_{21}(z,\zeta)=0$.
Making use of the resulting diagonal structure of $\mat P(z,\zeta)$ and the method of characteristics, the remaining kernel elements
\begin{equation}
    P_{ii}(z,\zeta) = \frac{1}{\lambda_i(\zeta)}p_{i}(\phi_i(z)-\phi_i(\zeta)), \qquad i=1,2
\end{equation}
are given by the solution of the Volterra integral equations%
\begin{subequations}
\label{eq:bs_kerneleqs_inteq}
    \begin{align}
    \label{eq:bs_kerneleqs_inteq1}
        \tint_0^{\tau} \vect e_1^\top \mat N(\psi_1(\tau-\sigma))\vect b p_1(\sigma)\,\dd\sigma + p_1(\tau) &= \vect e_1^\top \mat N(\psi_1(\tau)) \vect b \\
    \label{eq:bs_kerneleqs_inteq2}
        \tint_0^{\tau} \vect e_2^\top \mat N(\psi_2(\tau-\sigma))\vect b p_2(\sigma)\,\dd\sigma - q_0p_2(\tau) &= \vect e_2^\top \mat N(\psi_2(\tau)) \vect b
    \end{align}
\end{subequations}
with $\tau\in[0,\phi_1(1)]$ in \eqref{eq:bs_kerneleqs_inteq1} and $\tau\in[0,\phi_2(1)]$ in \eqref{eq:bs_kerneleqs_inteq2} that follow from \eqref{eq:bs_kerneleqs_bc0} after a transformation of the integration variable and a substitution of $\phi_1(z)$ (resp.\ $\phi_2(z)$) by $\tau$.

Because \eqref{eq:bs_kerneleqs} guarantees \eqref{eq:bs_pre_transportpde} and $\bar{\vect\varepsilon}(0,t)=\tilde{\vect\varepsilon}(0,t)$, the transformation \eqref{eq:bs_trafo_bs} maps \eqref{eq:bs_sys_decoupl} into
\begin{subequations}
\label{eq:bs_sys_bs}
    \begin{align}
    \label{eq:bs_sys_bs_ode}
        \dot{\vect e}_\xi(t) &= (\mat F+\vect b\vect k^\top)\vect e_\xi(t) + \vect b \bar\varepsilon_1(0,t) \\
    \label{eq:bs_sys_bs_bc0}
        \bar\varepsilon_2(0,t) &= q_0 \bar\varepsilon_1(0,t) \\
    \label{eq:bs_sys_bs_pde}
        \partial_t \bar{\vect\varepsilon}(z,t) &= \mat\Lambda(z) \partial_z \bar{\vect\varepsilon}(z,t) \\
    \label{eq:bs_sys_bs_bc1}
        \bar\varepsilon_1(1,t) &= \bar q_1\varepsilon_2(1,t) + \bar u(t) - \bar u_\soll(t) - \vect e_1^\top \mat N(1)\vect e_\xi(t) \nonumber\\
        &\hspace{2cm} - \tint_0^1 \vect e_1^\top\mat P(1,z)\tilde{\vect\varepsilon}(z,t)\,\dd z,
    \end{align}
\end{subequations}
with \glspl{IC} $\vect e_\xi(0)=\vect e_{\xi,0}$ and $\bar{\vect\varepsilon}(z,0)=\bar{\vect\varepsilon}_0(z)=\vect\varepsilon_0(z)-\mat N(z)\vect e_{\xi,0} - \tint_0^z \mat P(z,\zeta)[\vect\varepsilon_0(\zeta) - \mat N(\zeta)\vect e_{\xi,0}]\,\dd\zeta$ for \eqref{eq:bs_sys_bs} defined based on those of \eqref{eq:bs_error_sys}.
In view of the invertible transformations \eqref{eq:bs_trafo_decoupl} and \eqref{eq:bs_trafo_bs}, where the inverse map of \eqref{eq:bs_trafo_bs} and the reciprocity relation yielding $\mat P_\mathrm{I}(z,\zeta)\in\Rset^{2\times2}$ are defined analogous to \eqref{eq:pre_trafo_bs_inv} and \eqref{eq:reciprocity}, respectively, \eqref{eq:bs_error_sys} and \eqref{eq:bs_sys_bs} are equivalent.

\myvspace{.25cm}
\begin{lem}[Equivalence of \eqref{eq:bs_error_sys} and \eqref{eq:bs_sys_bs}]
\label{lem:strictfeedback}
    The form \eqref{eq:bs_sys_bs} is an equivalent representation of \eqref{eq:bs_error_sys}.
\end{lem}

Owing to the coupling structure of \eqref{eq:bs_sys_bs} and the stable ODE subsystem \eqref{eq:bs_sys_bs_ode}, a state feedback for $\bar u(t)$ only has to ensure the stability of the PDE subsystem \eqref{eq:bs_sys_bs_bc0}--\eqref{eq:bs_sys_bs_bc1} in order for the overall system to be stable.
With the PDE subsystem described in its most simple form by two cascaded transport equations, this choice is very easy (recall Figure~\ref{fig:normalform-overview}).
It is important to note that choosing an appropriate feedback would no longer be obvious if the ODE subsystem \eqref{eq:bs_sys_bs_ode} were not already stable, as a controller only compensates the potentially destabilizing terms at the actuated boundary \eqref{eq:bs_sys_bs_bc1}.

\begin{figure}
    \centering
    \includegraphics{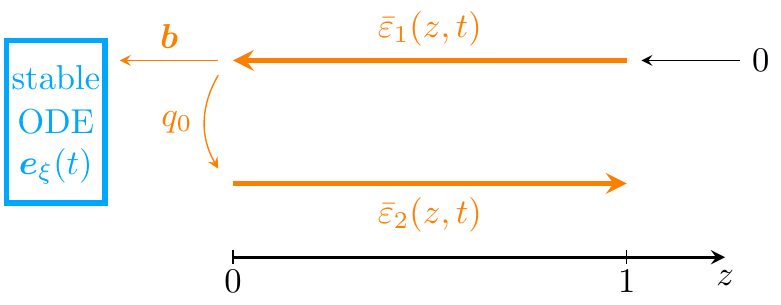}
    \caption{Visualization of the closed-loop system \eqref{eq:bs_sys_bs_ode}--\eqref{eq:bs_sys_bs_pde}, \eqref{eq:bs_sys_bs_bc1_closedloop} for $\bar q_{1,\mathrm{cl}}=0$.}
    \label{fig:targetsys_structure}
\end{figure}

For example, stability of the closed loop is ensured if the \gls{BC} \eqref{eq:bs_sys_bs_bc1} at $z=1$ takes the form
\begin{equation}
\label{eq:bs_sys_bs_bc1_closedloop}
    \bar\varepsilon_1(1,t) = \bar q_{1,\mathrm{cl}}\bar\varepsilon_2(1,t),
\end{equation}
with $\bar q_{1,\mathrm{cl}}$ such that $|q_0\bar q_{1,\mathrm{cl}}|<1$.
The exponential stability of the PDE subsystem and thus of \eqref{eq:bs_sys_bs_ode}--\eqref{eq:bs_sys_bs_pde} with \eqref{eq:bs_sys_bs_bc1_closedloop} is a classical result (see, e.g., \cite[Thm.~2.2]{Fridman2014} in the context of time-delay systems).
Although Remark~\ref{rem:bs_cl} addresses alternative closed-loop dynamics, most commonly $\bar q_{1,\mathrm{cl}}=0$ is chosen.
The corresponding coupling structure of the closed-loop system \eqref{eq:bs_sys_bs} with \eqref{eq:bs_sys_bs_bc1_closedloop} instead of \eqref{eq:bs_sys_bs_bc1} is sketched in Figure~\ref{fig:targetsys_structure}.
Therein, it is apparent that the input $\bar\varepsilon_1(0,t)$ of the stable ODE vanishes for $t>\tau_1$ if $\bar q_{1,\mathrm{cl}}=0$ and the PDE state $\bar{\vect\varepsilon}(z,t)$ itself is zero after the finite time $\tau_1+\tau_2$.
A comparison of \eqref{eq:bs_sys_bs_bc1} and \eqref{eq:bs_sys_bs_bc1_closedloop} then directly yields the backstepping-based tracking controller
\begin{align}
\label{eq:bs_controller}
    \bar u(t) &= \bar u_\soll(t) + (\bar q_{1,\mathrm{cl}}-\bar q_1)\varepsilon_2(1,t) \\
    &+ \tint_0^1 (\vect e_1^\top-\bar q_{1,\mathrm{cl}}\vect e_2^\top)\mat P(1,z)\vect\varepsilon(z,t)\,\dd z \nonumber\\
    &+ (\vect e_1^\top-\bar q_{1,\mathrm{cl}}\vect e_2^\top)\!\left[\mat N(1) - \tint_0^1 \mat P(1,z)\mat N(z)\,\dd z\right] \vect e_\xi(t), \nonumber
\end{align}
written as a feedback of the original error states $\vect e_\xi(t)$ and $\vect\varepsilon(z,t)$ of \eqref{eq:bs_error_sys} by use of the transformations \eqref{eq:bs_trafo_decoupl} and \eqref{eq:bs_trafo_bs}.

\begin{figure*}[t]
    \centering
    \scalebox{.97}{\includegraphics{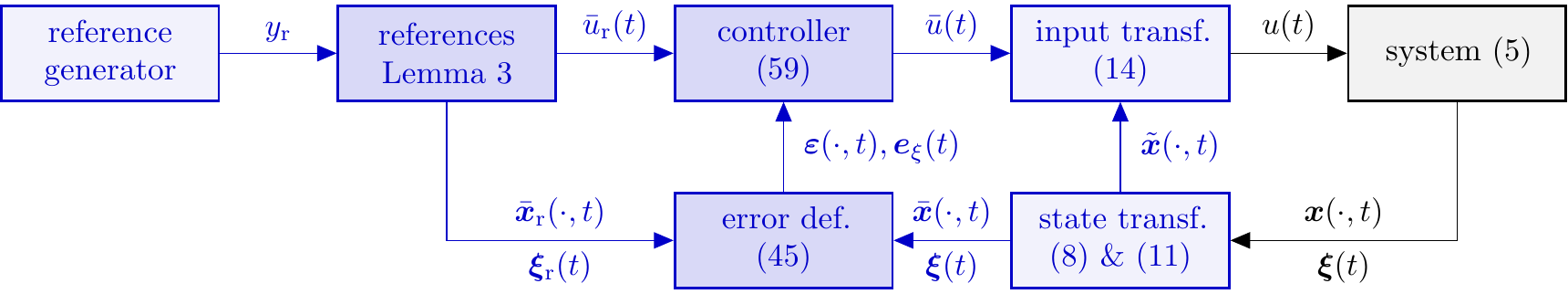}}
    \caption{Signal flow diagram of the closed-loop system using the backstepping controller \eqref{eq:bs_controller} to track a reference \eqref{eq:sys_reference} associated with $\bar{\vect x}_\soll(z,t)$, $\vect\xi_\soll(t)$ (and potentially obtained from \eqref{eq:sys_reference}) for the PDE-ODE system \eqref{eq:sys}.
    Blocks with a stronger saturation highlight the main design components and the differences to the flatness-based controller in Figure~\ref{fig:flat-signalflow}.}
    \label{fig:bs-signalflow}
\end{figure*}

The overall closed-loop system is visualized by the signal flow diagram in Figure~\ref{fig:bs-signalflow}.
Its stability is asserted in the following theorem.

\begin{thm}[Closed-loop stability]
\label{thm:bs}
    Let Assumptions \ref{enum:ass_ode} and \ref{enum:ass_pde} hold, with $\vect k$ such that $\mat F+\vect b\vect k^\top$ is Hurwitz and $\bar q_{1,\mathrm{cl}}$ such that $|q_0\bar q_{1,\mathrm{cl}}|<1$.
    Then, the state feedback controller \eqref{eq:bs_controller} exponentially stabilizes the system \eqref{eq:sys} along reference trajectories \eqref{eq:sys_reference}.
\end{thm}

Based on Lemma~\ref{lem:strictfeedback}, exponential stability of the closed-loop system \eqref{eq:bs_sys_bs_ode}--\eqref{eq:bs_sys_bs_pde}, \eqref{eq:bs_sys_bs_bc1_closedloop} implies convergence of $\vect\xi(t)$ towards $\vect\xi_\soll(t)$ and of $\bar{\vect x}(z,t)$ towards $\bar{\vect x}_\soll(z,t)$ pointwise in space (see  \cite[Thm.~4]{Deutscher2017WC} for the case $\bar q_{1,\mathrm{cl}}=0$).
This ensures the tracking of a reference \eqref{eq:sys_reference} for the original system \eqref{eq:sys} by Lemma \ref{lem:equiv_sys_pre} and Lemma \ref{lem:references}.

\begin{rem}[Choice of closed-loop dynamics]
\label{rem:bs_cl}
    Any choice of a controller that results in an exponentially stable closed-loop dynamics for the PDE subsystem is admissible.
    For example, an additional transformation of the form \eqref{eq:pre_trafo_hopfcole} may be used to introduce reaction terms in the transport PDE \eqref{eq:bs_sys_bs_pde} and to have the solutions on the characteristic curves decay, with the feedback again following from the \gls{BC} at $z=1$.
    Oftentimes, a backstepping design makes use of the additional assumption $|q_0\bar q_1|<1$ (e.g.\ \cite{Auriol2018AUT}), which allows to set $\bar q_{1,\mathrm{cl}}=\bar q_1$ in view of the stability condition $|q_0\bar q_{1,\mathrm{cl}}|<1$.
    In the context of time-delay systems, this choice is known to be robust w.r.t.\ small delays in the feedback loop that may arise in a controller implementation due to the cancellation of the boundary term (see coefficient of $\varepsilon_2(1,t)$ in \eqref{eq:bs_controller}).
\end{rem}

  \section{A comparison of both designs}
\label{sec:comparison}

The controller designs in Sections~\ref{sec:flatness} and \ref{sec:backstepping} are clearly based on normal forms (see Figure~\ref{fig:normalform-overview}).
Although they both share the idea of using state transformations to map a given system representation into a form from which a stabilizing feedback is easily inferred, they have quite a few differences, at least at first glance.

With an additional transport equation to incorporate the infinite-dimensional dynamics, the \gls{HCCF} \eqref{eq:flat_hccf} used in the flatness-based design is a direct generalization of the \gls{CCF} \eqref{eq:ccf_finite} to hyperbolic PDE-ODE systems \eqref{eq:sys}.
The feedback simply follows from replacing the system's characteristics by some desired one for the closed loop.
In contrast, the backstepping approach in Section~\ref{sec:backstepping} transfers the design of the same name to infinite-dimensional systems \eqref{eq:sys} in strict-feedback form.
The controller is derived by successively stabilizing the subordinate subsystems and in the process mapping the dynamics into a simpler form.
In the end, similar but not identical to the flatness-based design based on the \gls{HCCF} \eqref{eq:flat_hccf}, the choice of a desired \gls{BC} \eqref{eq:bs_sys_bs_bc1} (analogous to \eqref{eq:flat_hccf_bc}) for the closed-loop system directly yields a stabilizing feedback.
Due to the cascaded structure of \eqref{eq:bs_sys_bs}, this only assigns a stable dynamics to the PDE subsystem, with the ODE subsystem in closed loop already fixed by the choice of $\vect k$ in the backstepping transformation \eqref{eq:bs_trafo_decoupl}.

If the two transformations \eqref{eq:bs_trafo_decoupl} and \eqref{eq:bs_trafo_bs} involved in the backstepping design are combined into a single one, the result actually strongly resembles the structure \eqref{eq:flat_trafo_compact} of the integral transformation into \gls{HCCF}.
Both are essentially of Volterra type with an additional dependence on the ODE state (although one uses the tracking errors \eqref{eq:bs_error_def}).
The main difference between both transformations is that \eqref{eq:flat_trafo_compact_ode} allows to map the ODE subsystem \eqref{eq:bs_sys_bs_ode} into the chain of integrators in \eqref{eq:flat_hccf_etai}, which is fundamental in the \gls{HCCF}.
On the other hand, Section~\ref{sec:backstepping} does without a transformation of the ODE state, as the design only relies on the strict-feedback form of the system, and choosing a stabilizing virtual feedback \eqref{eq:bs_virtual_feedback} is easy enough without a special structure of $\mat F$.
At this point, it should be emphasized that the preliminary transformations in Section~\ref{sec:prelim} are not necessary for either of the controller designs.
Rather, they simplify solving the Cauchy problem w.r.t.\ $z$ (see \eqref{eq:flat_sol_pde}) for the derivation of the flatness-based parametrization and allow for an explicit solution of the initial value problem \eqref{eq:bs_decouple_ivp} in the backstepping approach.
This further underlines the advantage of a multi-step design over a single transformation.

Using multiple state transformations for the derivation of a controller also simplifies the numerical implementation.
The backstepping controller \eqref{eq:bs_controller} requires the solutions $\mat N(z)$ and $\mat P(z,\zeta)$ of the initial value problem \eqref{eq:bs_decouple_ivp} and the kernel equations \eqref{eq:bs_kerneleqs}, respectively.
As both are standard equations encountered in multi-step designs for different type of systems (see, e.g., \cite{Deutscher2018AUT,Gehring2021MTNS}), a toolbox (e.g.~\cite{coni}) with appropriate modular functions can be used.
In contrast, a one-step design (e.g.~\cite{DiMeglio2018aut}) yields bilaterally coupled decoupling and kernel equations that are hard to solve.
More importantly, such a design paradigm leads to different kernel equations for every new system structure, which impedes the implementation of their solution.

While the complexity of the backstepping design is mainly hidden in the explicit solution of \eqref{eq:bs_decouple_ivp} and \eqref{eq:bs_kerneleqs}, which is not discussed here in detail, the flatness-based perspective uses the explicit system solutions and, thus, may mistakenly be perceived as more complicated.
However, the calculations for the \gls{HCCF} state transformation in Appendix~\ref{sec:app_hccf_trafo} basically correspond to the explicit solution of \eqref{eq:bs_decouple_ivp} and \eqref{eq:bs_kerneleqs} in the backstepping design.
The temporal integrals in Appendix~\ref{sec:app_hccf_trafo} are related to the spatial ones involved in the solution of \eqref{eq:bs_decouple_ivp} and \eqref{eq:bs_kerneleqs} via the characteristic curves of \eqref{eq:sys_transp_pde}.
Still, recall that the \gls{HCCF} and thus the corresponding state transformation is not required for the implementation of the flatness-based feedback.

Ultimately, both controllers \eqref{eq:flat_controller} and \eqref{eq:bs_controller} guarantee exponential stability of the closed-loop system in an appropriate sense (see Theorems~\ref{thm:flatness} and \ref{thm:bs}).
In fact, both state feedbacks are identical if $\gamma$ and $\kappa_i$, $i=0,\dots,n-1$, in the flatness-based design and $\bar q_{1,\mathrm{cl}}$ and $\vect k$ for the backstepping approach are chosen such that
\begin{subequations}
\label{eq:compare_identical}
    \begin{align}
        \gamma &= -q_0\bar q_{1,\mathrm{cl}} \\
        s^n + \sum_{i=0}^{n-1} \kappa_i s^i &= \det(s\mat I-\mat F-\vect b\vect k^\top),
    \end{align}
\end{subequations}
which also reflects Assumptions~\ref{enum:ass_pde} and \ref{enum:ass_ode}.
This follows from a comparison of the closed-loop systems \eqref{eq:flat_closedloop} and \eqref{eq:bs_sys_bs}, with \eqref{eq:bs_sys_bs_bc1_closedloop} instead of \eqref{eq:bs_sys_bs_bc1} in the latter.
Correspondences like \eqref{eq:compare_identical} may no longer be obvious or even exist if a different dynamics is chosen for the closed loop in either Section~\ref{sec:flatness} or \ref{sec:backstepping} (recall Remarks~\ref{rem:flat_cl} and \ref{rem:bs_cl}).
An equivalence analysis related to \eqref{eq:compare_identical} can also be found in \cite{Irscheid2022chapter} in the context of the solution-based control of PDE-ODE systems with a nonlinear ODE subsystem.
The equivalence of the closed-loop systems in Figures~\ref{fig:flat-signalflow} and \ref{fig:bs-signalflow} for appropriate design parameters is also discussed in the following example.
  \section{Example: Heavy rope with load}
\label{sec:example}

The controller designs in Sections~\ref{sec:flatness} and \ref{sec:backstepping} are applied to the classical, linear model of a heavy rope with a load (e.g.~\cite{Petit2001siam}) that also serves as the example for the backstepping approach in \cite{Deutscher2017WC}.
Numerical results illustrate the control performance for different choices of the closed-loop dynamics.
In that context, further details on the implementation of both state feedback controllers are provided.


\subsection{System specification}

\begin{figure}[t]
    \centering
    \includegraphics[width=\columnwidth]{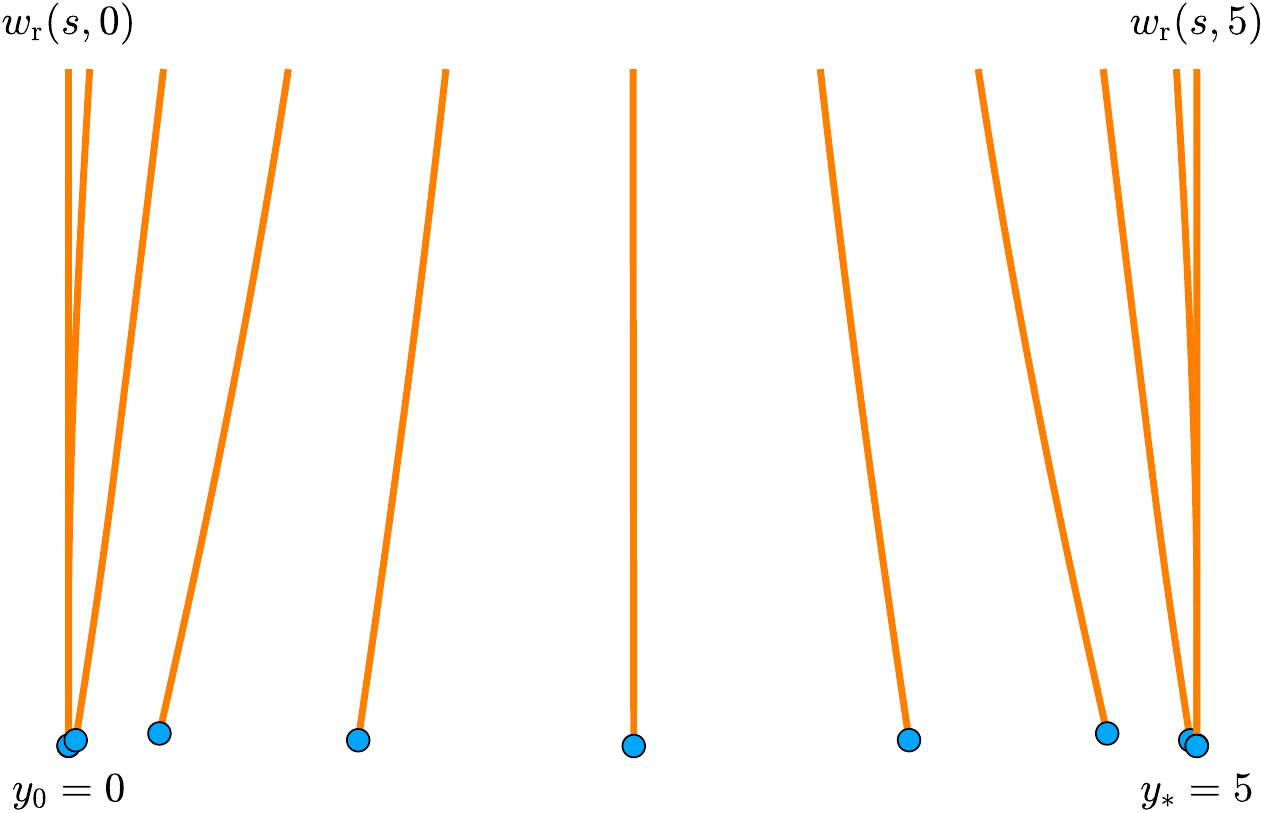}
    \caption{Visualization of the desired transition between the two steady states $w_\soll(s,0)=0$ and $w_\soll(s,5)=5$ of the heavy rope with load.
    The transition is specified by the reference $y_\soll(t)$ for the load position, which is a flat output of the system.}
    \label{fig:ex-animation}
\end{figure}

Consider the planar motion of a homogeneous rope of length $\ell$ with constant cross-sectional area and line density $\rho$ in the earth's gravitational field, with acceleration $g$.
Using the curvilinear coordinate $s\in[0,\ell]$ along the rope, $w(s,t)$ denotes the horizontal displacement of the rope relative to a fixed vertical reference at time $t$.
As sketched in the context of Figure~\ref{fig:ex-animation}, a point load of mass $m$ is attached to the lower end of the rope at $s=0$.
The velocity at $s=\ell$ serves as control input.
Using small-angle approximations, a momentum balance yields the model
\begin{subequations}
\label{eq:ex-rope}
    \begin{align}
    \label{eq:ex-rope-pde}
        \rho\partial_t^2 w(s,t) &= \partial_{s}\big(g(\rho s + m) \partial_{s} w(s,t)\big), \quad s\in(0,\ell) \\
    \label{eq:ex-rope_z0}
        \partial_t^2 w(0,t) &= g\partial_{s} w(0,t) \\
    \label{eq:ex-rope_z1}
        \partial_t w(\ell,t) &= u(t)
    \end{align}
\end{subequations}
for the motion of a heavy rope with a load.
In order to rewrite \eqref{eq:ex-rope} in the form \eqref{eq:sys}, first, the spatial domain $[0,\ell]$ is normalized by setting $z=\frac{s}{\ell}\in[0,1]$.
Then, in view of the dynamical \gls{BC} \eqref{eq:ex-rope_z0}, choosing states
\begin{subequations}
    \begin{align}
        \vect x(z,t) &= \frac{1}{2}
        \begin{bmatrix}
             \lambda(z) & 1 \\
            -\lambda(z) & 1 
        \end{bmatrix}
        \begin{bmatrix}
            \partial_{s} w(\ell z,t) \\
            \partial_t w(\ell z,t)
        \end{bmatrix}, \quad z\in[0,1] \\
        \vect\xi(t) &= 
        \begin{bmatrix}
            w(0,t) \\
            \partial_t w(0,t)
        \end{bmatrix}
    \end{align}
\end{subequations}
with
\begin{equation}
  \lambda(z) = \sqrt{\frac{g}{\rho}(\rho\ell z+m)}, \qquad z\in[0,1]
\end{equation}
and $n=2$ due to $\vect\xi(t)\in\Rset^2$ yields \eqref{eq:sys}.
Therein, 
\begin{equation}
    \mat F = 
    \begin{bmatrix}
        0 & 1 \\[.5ex]
        0 & -\frac{g}{\lambda(0)}
    \end{bmatrix}, \qquad
    \vect b = \begin{bmatrix} 0 \\[.5ex] \frac{2g}{\lambda(0)} \end{bmatrix}, \qquad
    \vect c = \begin{bmatrix} 0 \\[.5ex] 1 \end{bmatrix},
\end{equation}
$q_0=q_1=-1$ as well as $\lambda_1(z)=\lambda_2(z)=\frac{\lambda(z)}{\ell}$ and $A_{i1}(z)=-A_{i2}(z)=\frac{\lambda^\prime(z)}{2\ell}$, $i=1,2$.
It is easily verified that this PDE-ODE in strict-feedback form satisfies Assumptions~\ref{enum:ass_ode} and \ref{enum:ass_pde}.
Moreover, it is shown in \cite{Petit2001siam} that the heavy rope with a point mass admits the flat output $y(t)=w(0,t)=[1,0]\vect\xi(t)$ (cf.~\eqref{eq:flat_output}).

Inspired by an experimental setup, the parameters in \eqref{eq:ex-rope} are chosen as $\rho=0.3~\unitfrac{kg}{m}$, $\ell=3~\unit{m}$, $g=9.81~\unitfrac{kg}{ms^2}$ and $m=0.25~\unit{kg}$.
This implies delays $\tau_1=\tau_2\approx0.67~\unit{s}$ by \eqref{eq:transporttimes} as well as eigenvalues $0~\unitfrac{1}{s}$ and $-3.43~\unitfrac{1}{s}$ of $\mat F$.
Consequently, the uncontrolled ODE is not asymptotically stable.
In the following, all numerical values are given in appropriate SI units, which are omitted to simplify the presentation.

\subsection{Tracking problem}

The objective is to horizontally move the rope from one rest position to another (see Figure~\ref{fig:ex-animation}).
For that, a reference $y_\soll(t)$ for the flat output $y(t)=\xi_1(t)$, i.e. the position of the load, is specified according to \eqref{eq:ysoll}, with
\begin{subequations}
    \begin{align}
        c_0 &= y_0, & c_1 &= 0, & c_2 &= 0 \\
        c_3 &= 10(y_*-y_0), & c_4 &= -15(y_*-y_0), & c_5 &= 6(y_*-y_0)
    \end{align}
\end{subequations}
for the polynomial of degree $5$ in \eqref{eq:ysoll_polynomial} due to $n=2$.
More precisely, the load is transitioned between the initial position $y_\soll(t_0)=y_0=0$ at time $t_0=\tau_1$ and the final one $y_\soll(t_*)=y_*=5$ at $t_*=5-\tau_2$.
Note that the time interval $[t_0,t_*]\approx[0.67,4.33]$ is chosen such that the feedforward controller \eqref{eq:flat_feedforward} achieves the transition over $[0,5]$ (recall Figure~\ref{fig:reference}).
Applying the designs in Sections~\ref{sec:flatness} and \ref{sec:backstepping} yields two tracking controllers.

\begin{figure}[t]
    \centering
    \includegraphics[width=\columnwidth]{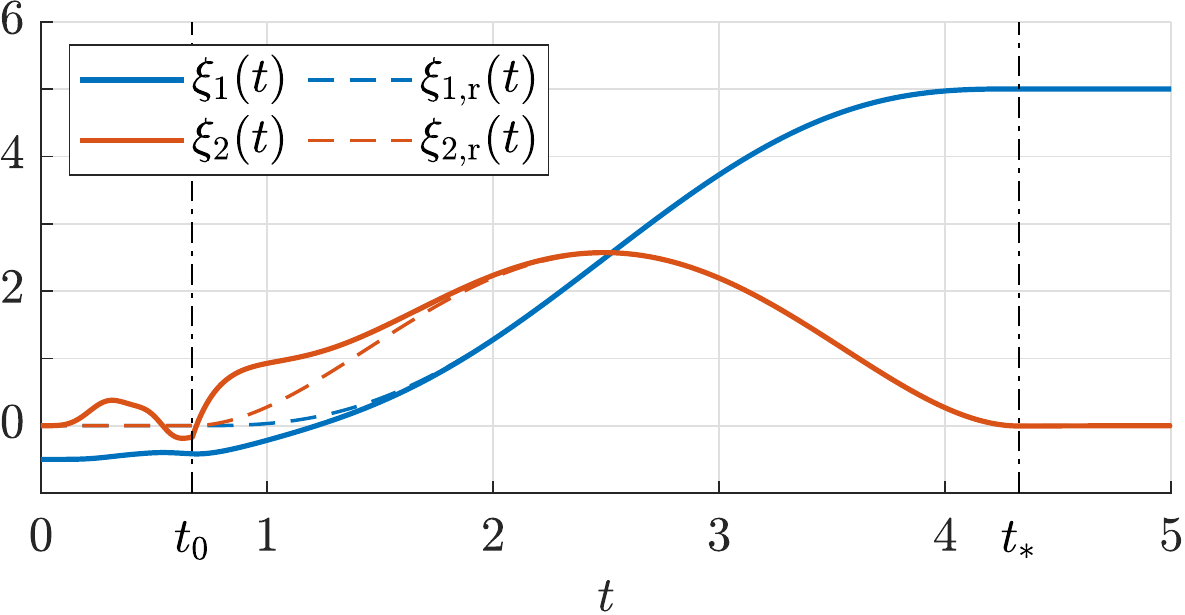}
    \caption{Case $\gamma=0$: The components of the ODE state $\vect\xi(z,t)$ converge to their respective reference for $t>\tau_1\approx0.67$.}
    \label{fig:ex-ode-state}
\end{figure}

\subsection{Controller design and implementation}

Three different sets of controller parameters are considered, each such that the flatness-based design and the backstepping method yield the same feedback (see \eqref{eq:compare_identical}).
For the flatness-based controller, $\kappa_0=20$, $\kappa_1=9$ and $\gamma\in\{-0.3,0,0.3\}$ are chosen in the error dynamics \eqref{eq:flat_closedloop}. 
By \eqref{eq:compare_identical}, this fixes $\bar q_{1,\mathrm{cl}}=-\frac{\gamma}{q_0}=\gamma\in\{-0.3,0,0.3\}$ in \eqref{eq:bs_sys_bs_bc1} and the vector $\vect k$ that determines the solution of the initial value problem \eqref{eq:bs_decouple_ivp} in the backstepping design.
The division by $q_0$ in the determination of $\bar q_{1,\mathrm{cl}}$ further emphasizes the necessity of Assumption~\ref{enum:ass_pde} for choosing arbitrary closed-loop dynamics when using backstepping.

The implementation of the controllers follows the signal flow diagrams in Figures~\ref{fig:flat-signalflow} and \ref{fig:bs-signalflow}.
Both $y_\soll(t)$ and the references in \eqref{eq:sys_reference} defined by it (see Lemmas~\ref{lem:references} and \ref{lem:equiv_sys_pre}) are calculated offline.
Note that the determination of the references required for the backstepping controller is comparatively extensive.
In general, all integrals appearing in the implementation of the two controllers are discretized by the trapezoidal rule.

The flatness-based feedback \eqref{eq:flat_controller} of the \gls{HCCF} state makes use of the transformation \eqref{eq:flat_trafo_compact}.
The functions therein can be determined either analytically offline, which may be cumbersome, or simply numerically online.
In any case, there is no need for an online solution of Volterra integral equations.
As seen in Figure~\ref{fig:flat-signalflow}, moreover, the design uses the kernel $\mat K(z,\zeta)$ of the preliminary transformation \eqref{eq:pre_trafo_bs} to obtain a feedback of the states $\vect\xi(t)$ and $\vect x(z,t)$ of \eqref{eq:sys}.
In fact, both controllers require $\mat K(z,\zeta)$.
A numerical solution of the associated kernel equations \eqref{eq:pre_kerneleqs} is calculated offline on a fixed spatial grid with the help of the toolbox \cite{coni}.
Similarly, the initial value problem \eqref{eq:bs_decouple_ivp} and the kernel equations \eqref{eq:bs_kerneleqs} are solved offline using the same spatial grid.
The numerical solutions for $\mat K(z,\zeta)$, $\mat N(z)$ and $\mat P(z,\zeta)$ form the basis for the implementation of the backstepping controller \eqref{eq:bs_controller} in Figure~\ref{fig:bs-signalflow}.




\begin{figure}[t]
    \centering
    \includegraphics[width=\columnwidth]{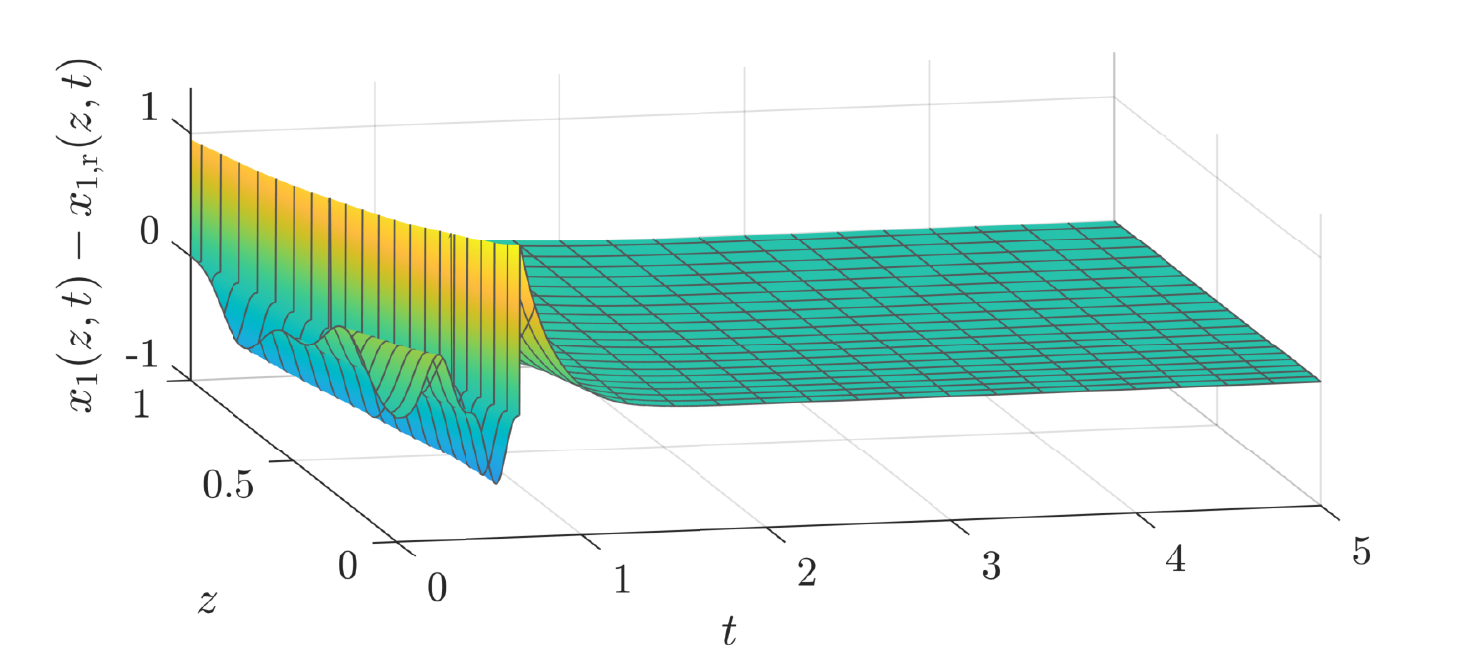}
    \\
    \includegraphics[width=\columnwidth]{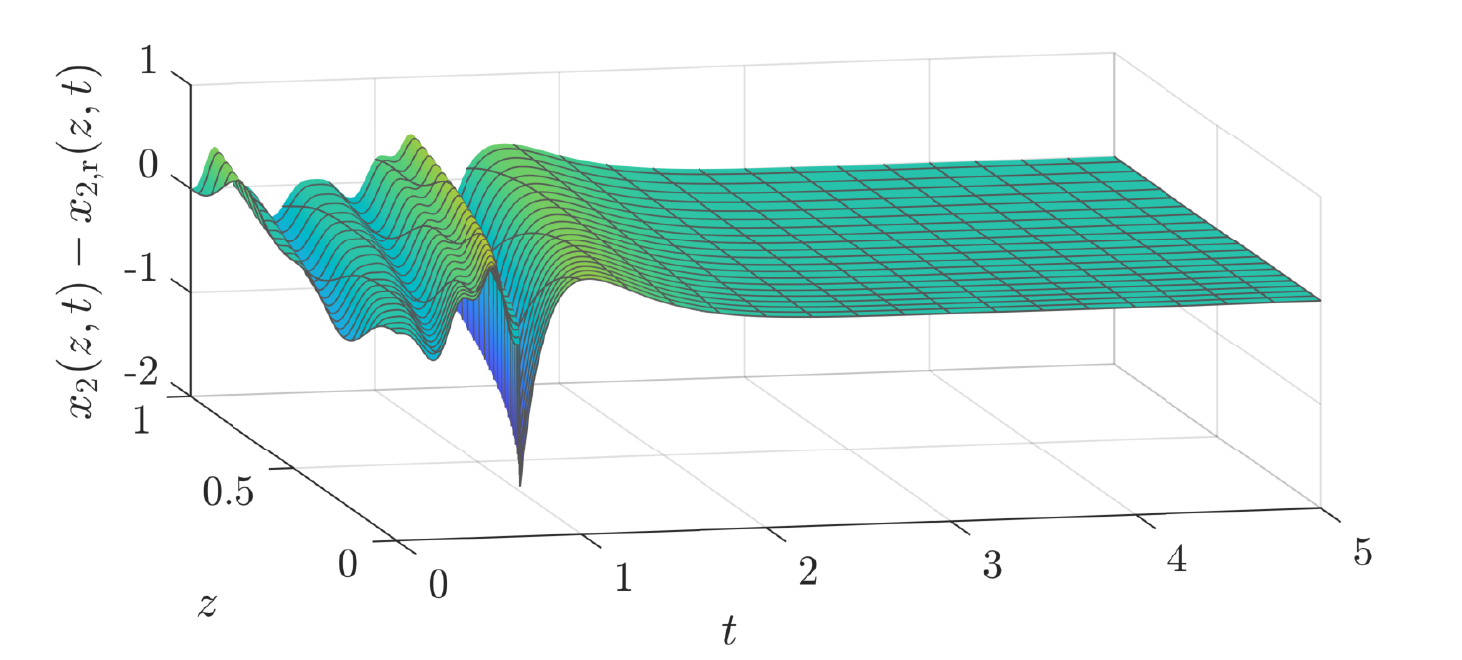}
    \caption{Case $\gamma=0$: The components of the PDE error state $\vect x(z,t)-\vect x_\soll(z,t)$ converge to zero for $t>\tau_1\approx0.67$.}
    \label{fig:ex-pde-state}
\end{figure}

\subsection{Simulation results}

The implementation of system \eqref{eq:sys} makes use of the method of characteristics, with the explicit Euler method used for numerical integration.
As such, the step size $2.5\cdot 10^{-3}$ chosen for the time $t$ implies the spatial grid of the distributed state $\vect x(z,t)$.
The \glspl{IC}
\begin{equation}
    \vect\xi_0 = \begin{bmatrix} -\frac{1}{2} \\[.5ex] 0 \end{bmatrix}, \qquad \vect x_0(z) = \frac{2}{5}\sin^3(2\pi z)\begin{bmatrix} 1 \\[.5ex] -1 \end{bmatrix}
\end{equation}
of the system are such that the rope is neither in steady state nor is the load at the position used as the starting position for the transition.
The flatness-based and the backstepping-based controllers are implemented as feedbacks of the states $\vect\xi(t)$ and $\vect x(z,t)$ of \eqref{eq:sys} (cf.\ Figure~\ref{fig:flat-signalflow} and \ref{fig:bs-signalflow}).
Simulation results confirm that both controllers yield the same closed-loop behavior in view of \eqref{eq:compare_identical}, apart from very minor deviations that are solely due to numerical issues associtated with the chosen step size.
Therefore, the following results reflect both designs equally, even though only values for $\gamma$ are referenced as $\gamma=\bar q_{1,\mathrm{cl}}$.

For $\gamma=0$, Figures~\ref{fig:ex-ode-state} and \ref{fig:ex-pde-state} show that the states $\vect\xi(t)$ and $\vect x(z,t)$ of \eqref{eq:sys} converge to their respective references $\vect\xi_\soll(t)$ and $\vect x_\soll(z,t)$, with initial errors compensated and almost no error left for $t>2.5$.
In particular, the convergence is exponential in time for $t>\tau_1\approx0.67$ because of the closed-loop dynamics in \eqref{eq:flat_closedloop} with $\gamma=0$ as well as those in \eqref{eq:bs_sys_bs} with \eqref{eq:bs_sys_bs_bc1_closedloop} instead of \eqref{eq:bs_sys_bs_bc1}.

\begin{figure}[t]
    \centering
    \includegraphics[width=\columnwidth]{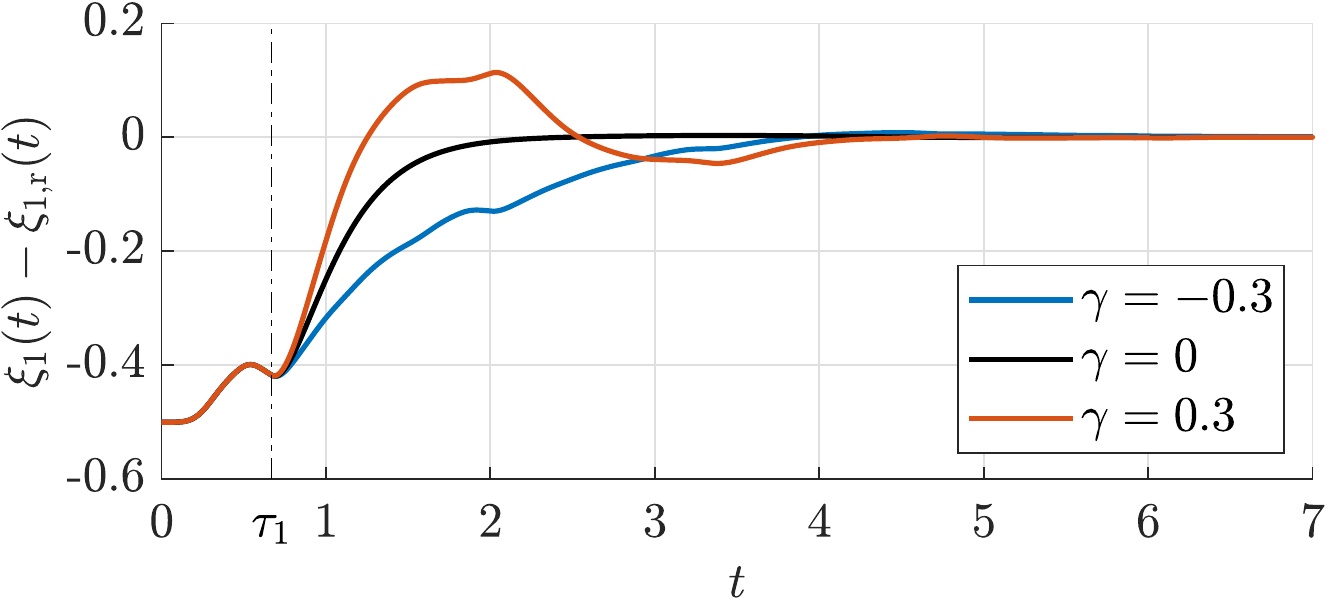}
    \caption{Deviation of the ODE state component $\xi_1(t)$, which is a flat output, from its reference for different values of $\gamma$. }
    \label{fig:ex-gamma}
\end{figure}

In Figure~\ref{fig:ex-gamma}, the control performance for different values of $\gamma$ is compared based on the error $\xi_1(t)-\xi_{1,\soll}(t)$.
This is especially interesting as $\xi_1(t)$ is a flat output of the system and the convergence of a flat output to its corresponding reference implies the same for all system variables.
Looking at Figure~\ref{fig:ex-gamma}, it becomes apparent that the evolution of $\xi_1(t)$ for $t\in[0,\tau_1]$ is independent of $\gamma$.
This is unsurprising as the control action at $z=1$ takes the transport time $\tau_1$ to affect the ODE at $z=0$.
Overall, the case $\gamma=0$, which corresponds to the results in Figures~\ref{fig:ex-ode-state} and \ref{fig:ex-pde-state}, demonstrates the fastest error convergence.
Instead, the smallest rise time is observed for $\gamma=0.3$, which is due to the alternating behavior of the error that is evident based on the associated closed-loop dynamics.
For $\gamma=-0.3$ and $\gamma=0.3$, it takes the same amount of time for the error to be zero, about twice as long as in the case $\gamma=0$.
Still, a non-zero value for $\gamma$ may be advantageous when looking at the control effort.
The root mean square $u_\mathrm{rms}$ for $u(t)-u_\soll(t)$ is $u_\mathrm{rms}\approx0.12$ for $\gamma=-0.3$, $u_\mathrm{rms}\approx0.16$ for $\gamma=0$ and $u_\mathrm{rms}\approx0.21$ for $\gamma=0.3$, i.e., it increases with larger values of $\gamma$.
The upside of $\gamma=-0.3$ is apparent from a comparison of \eqref{eq:flat_baru_hccf} and \eqref{eq:flat_closedloop_rewrite} or alternatively of \eqref{eq:bs_sys_bs_bc1} and \eqref{eq:bs_sys_bs_bc1_closedloop} in open and closed loop, respectively.
With $-q_0\bar q_1=-1$ the counterpart of $\gamma$ in open loop, a negative choice for $\gamma$ essentially means that $u(t)$ does less compensation of the backward-traveling wave (see local terms $\eta_{n+1}(-\tau_2,t)$ in \eqref{eq:flat_controller} and $\varepsilon_2(1,t)$ in \eqref{eq:bs_controller}), thus requiring less control effort.
Note that the delay-robust feedback mentioned in Remark~\ref{rem:bs_cl} is not possible here because $q_0\bar q_1=1$.

  \section{Concluding remarks}
\label{sec:conclusion}

Both control strategies presented rely on normal forms and state transformations to determine a stabilizing (static) state feedback.
The flatness-based design in Section~\ref{sec:flatness} maps the PDE-ODE system \eqref{eq:sys} into the \gls{HCCF} to find a controller.
In contrast, the backstepping technique in Section~\ref{sec:backstepping} starts off with the strict-feedback form and recursively stabilizes the two subsystems of \eqref{eq:sys}.
As such, both approaches directly generalize respective designs in the finite-dimensional setting.

The controllers \eqref{eq:flat_controller} and \eqref{eq:bs_controller} can be augmented to output feedback tracking controllers.
For that, usually a state observer is designed that provides estimates of the PDE and ODE states based on a boundary measurement.
Analogous to the controllability assumptions \ref{enum:ass_ode} and \ref{enum:ass_pde}, this presumes observability or at least detectability of the corresponding PDE-ODE system.
In the context of an observer design using normal forms, it is shown in \cite{Woittennek2013nDs, Woittennek2012at} that the hyperbolic observer canonical form is dual to the \gls{HCCF} and allows for a very simple observer design.
The observer is derived from an input-output equation that takes the role of the flatness-based input parametrization \eqref{eq:flat_baru_hccf} as an equivalent representation of the system dynamics.
A flatness-based output feedback tracking control design for the example of a pneumatic \gls{DPS} can be found in \cite{Gehring2022letters}.
Backstepping controllers have been designed using different collocated, anti-collocated or pointwise in-domain measurements.
While the state observation based on measuring $x_2(1,t)$ can be shown to be simply dual to the state feedback design for \eqref{eq:sys}, the observers in \cite{Deutscher2019ijc} and \cite{DiMeglio2020aut} are more involved.
In any case, existing backstepping results rarely make use of a normal form or system structure to facilitate the observer design, with observability assumptions at time more related to a specific approach rather than a system property.


On the other hand, it is well known, that the design of backstepping controllers is not limited to linear hyperbolic PDE-ODE systems \eqref{eq:sys} with a scalar input.
Rather, extensions exist towards multi-input multi-output ODE-PDE-ODE systems with a general heterodirectional hyperbolic subsystem of arbitrary order and an additional ODE subsystem at the actuated boundary (e.g.~\cite{Deutscher2018AUT,DiMeglio2020aut,Gehring2021MTNS}), interconnected systems that are underactuated (e.g.~\cite{Auriol2021ecc}) or ODE-PDE-ODE systems of parabolic type (e.g.~\cite{Deutscher2021TAC,Gehring2021MTNS}).
Almost all of the cited references exploit certain system structures, e.g.\ a strict-feedback form, to allow for a simplified, recursive design.
Still, stabilizability and controllability assumptions differ, not only depending on whether stabilization of an equilibrium is sufficient or a trajectory is to be tracked.
This further emphasizes the importance of a thorough system analysis ahead of a controller design, e.g.\ by checking for the property of flatness.
Although extensions in the context of the flatness-based approach for \glspl{DPS} are sparse, in \cite{Woittennek2022CPDE}, the design in Section~\ref{sec:flatness} is generalized to PDE-ODE systems with a nonlinear ODE subsystem.
In fact, beyond what is already said in Section~\ref{sec:comparison}, there seems to be a fluid transition between the backstepping design and the flatness-based perspective.
Both are combined in \cite{Irscheid2023AUT} to solve the problem of output regulation for PDE-ODE systems with a nonlinear ODE subsystem.
In general, there appears to be a lot of untapped potential in the systematic design of state feedback controllers and observers using normal forms for \glspl{DPS}.

\appendix

  \section{HCCF state transformation}
\label{sec:app_hccf_trafo}

In what follows, the mapping between the state \eqref{eq:flat_hccf_state} of the \gls{HCCF} \eqref{eq:flat_hccf} and the states $\bar{\vect x}(z,t)$ and $\vect\xi(t)$ of \eqref{eq:sys_transp} is derived.
For that, \eqref{eq:flat_parametrization_z0_xi} reveals that
\begin{equation}
\label{eq:flat_HCCF_trafo_ODE}
    \vect\eta(t) = \mat T_c \vect\xi(t-\tau_2),
\end{equation}
the finite-dimensional part of the \gls{HCCF} state, corresponds to the delayed ODE state of \eqref{eq:sys_transp}.
As a relation with $\vect\xi(t)$ is sought, using the solution of the ODE \eqref{eq:sys_transp_ode} backwards in time, the delayed state is expressed by
\begin{multline}
\label{eq:flat_ODE_sol}
    \vect\xi(t-\phi_2(z)) = \rme^{-\mat F\phi_2(z)}\vect\xi(t) \\
    - \tint_0^{\phi_2(z)} \rme^{-\mat F(\phi_2(z)-\tau)}\vect b\bar x_1(0,t-\tau)\,\dd\tau
\end{multline}
for $z\in[0,1]$, which in turn depends on delayed values $\bar x_1(0,t-\tau)$, $\tau\in[0,\phi_2(z)]$.
It had previously been observed from the solution \eqref{eq:flat_sol_pde} of the Cauchy problem that predictions of the boundary value $\bar x_1(0,t)$ relate to the state component $\bar x_1(z,t)$ and delays of $\bar x_2(0,t)$ to $\bar x_2(z,t)$.
Based on that, solving the \gls{BC} \eqref{eq:sys_transp_bc0} for $\bar x_1(0,t)$, which is only possible because of the assumption $q_0\neq0$ (see \ref{enum:ass_pde}), and substituting the result in a slightly rewritten form of \eqref{eq:flat_sol_pde1} gives the delayed boundary value
\begin{align}
\label{eq:flat_x10_delayed}
    \bar x_1(0,t-\tau) &= \frac{1}{q_0}\big(\bar x_2(\psi_2(\tau),t) - \vect c^\top\vect\xi(t-\tau)\big) \\
    &\qquad - \frac{1}{q_0}\tint_0^\tau \vect e_2^\top \mat C(\psi_2(\sigma))\vect\xi(t-\tau+\sigma)\,\dd\sigma \nonumber
\end{align}
for $\tau\in[0,\phi_2(z)]$ in terms of the PDE state at $t$ and the delay ODE state.
In the end, the combination of \eqref{eq:flat_ODE_sol} and \eqref{eq:flat_x10_delayed} as well as a change of the order of integration yields the Volterra integral equation
\begin{align}
\label{eq:flat_inteq2}
    & \vect\xi(t-\phi_2(z)) - \tint_0^{\phi_2(z)} \bar{\mat C}_2(\phi_2(z)-\tau)\vect\xi(t-\tau)\,\dd\tau \\
    &\hspace{.2cm} = \rme^{-\mat F\phi_2(z)}\vect\xi(t) - \tint_0^{\phi_2(z)} \rme^{-\mat F(\phi_2(z)-\tau)}\frac{\vect b}{q_0}\bar x_2(\psi_2(\tau),t)\,\dd\tau \nonumber
\end{align}
for $\vect\xi(t-\phi_2(z))$ with $z\in[0,1]$, where $t$ taking the role of a parameter.
The convolution kernel
\begin{equation}
    \bar{\mat C}_2(\tau) = \rme^{-\mat F\tau}\frac{\vect b\vect c^\top}{q_0} + \tint_0^\tau \rme^{-\mat F(\tau-\sigma)}\frac{\vect b}{q_0}\vect e_2^\top\mat C(\psi_2(\sigma))\,\dd\sigma
\end{equation}
is defined for $\tau\in[0,\phi_2(z)]$.
The existence of a unique solution of \eqref{eq:flat_inteq2} is shown, e.g., in \cite[Thm.~3.11]{Linz1987} and allows to express the delayed state $\vect\xi(t-\phi_2(z))$ for all $z\in[0,1]$ in terms of the state at time $t$.
Thus, together with \eqref{eq:flat_HCCF_trafo_ODE}, the finite-dimensional part $\vect\eta(t)$ of the \gls{HCCF} follows from $\vect\xi(t)$ and the profile $\bar x_2(z,t)$.

In order to calculate $\eta_{n+1}(\tau,t)=y^{(n)}(t+\tau)$ based on $\vect\xi(t)$ and $\bar{\vect x}(z,t)$, first, apply a time shift $t\mapsto t+\tau$ to \eqref{eq:flat_parametrization_z0_x10} to get
\begin{equation}
\label{eq:flat_HCCF_trafo_PDE}
    \eta_{n+1}(\tau,t) = \bar x_1(0,t+\tau) + \vect e_n^\top \mat T_c \mat F \vect\xi(t+\tau).
\end{equation}
While $\tau\in[-\tau_2,\tau_1]$ for the distributed state \eqref{eq:flat_hccf_state_pde} of the \gls{HCCF}, for $\tau\in[-\tau_2,0]$, the delayed terms on the right-hand side of \eqref{eq:flat_HCCF_trafo_PDE} are already expressed in terms of $\vect\xi(t)$ and $\bar{\vect x}(z,t)$ (see \eqref{eq:flat_inteq2} and \eqref{eq:flat_x10_delayed}).
For $\tau\in[0,\tau_1]$, future values $\bar x_1(0,t+\tau)$ and $\vect\xi(t+\tau)$ are required in \eqref{eq:flat_HCCF_trafo_PDE}.
For that, similarly to the previous calculation of delayed values, the ODE \eqref{eq:sys_transp_ode} is solved forwards in time, while the shifted boundary values
\begin{equation}
\label{eq:flat_x10_predicted}
    \bar x_1(0,t+\tau) = \bar x_1(\psi_1(\tau),t) + \tint_0^\tau \vect e_1^\top \mat C(\psi_1(\sigma))\vect\xi(t+\tau-\sigma)\,\dd\sigma
\end{equation}
with $\tau\in[0,\phi_1(z)]$ follow from \eqref{eq:flat_sol_pde1}.
In the end, together with \eqref{eq:flat_x10_predicted}, the unique solution (see again \cite[Thm.~3.11]{Linz1987}) of the resulting Volterra integral equation
\begin{align}
\label{eq:flat_inteq1}
    & \vect\xi(t+\phi_1(z)) - \tint_0^{\phi_1(z)} \bar{\mat C}_1(\psi_1(z)-\tau)\vect\xi(t+\tau)\,\dd\tau \\
    &\hspace{.2cm} = \rme^{\mat F\phi_1(z)}\vect\xi(t) + \tint_0^{\phi_1(z)} \rme^{\mat F(\phi_1(z)-\tau)}\vect b\bar x_1(\psi_1(\tau),t)\,\dd\tau \nonumber
\end{align}
for $z\in[0,1]$, with the kernel
\begin{equation}
    \bar{\mat C}_1(\tau) = \tint_0^\tau \rme^{\mat F(\tau-\sigma)}\vect b\vect e_1^\top\mat C(\psi_1(\sigma))\,\dd\sigma,
\end{equation}
yields the right-hand side of \eqref{eq:flat_HCCF_trafo_PDE} for $\tau\in[0,\tau_1]$.

Based on the previous calculations, the transformation from $\vect\xi(t)$, $\bar{\vect x}(z,t)$ into $\vect\eta(t)$, $\eta_{n+1}(\tau,t)$ takes the form
\begin{subequations}
\label{eq:flat_trafo_compact}
    \begin{align}
    \label{eq:flat_trafo_compact_ode}
        \vect\eta(t) &= \mat G_0\vect\xi(t) - \int_0^1 \vect g_0(z) \bar x_2(z,t)\,\dd z \\
    \label{eq:flat_trafo_compact_pdep}
        \eta_{n+1}(\tau,t) &= \bar x_1(\psi_1(\tau),t) - \vect g_1^\top(\tau)\vect\xi(t) \\
        &\hspace{1cm} - \int_0^{\psi_1(\tau)} g_1(\tau,z)\bar x_1(z,t)\,\dd z, && \tau\in[0,\tau_1] \nonumber \\
    \label{eq:flat_trafo_compact_pdem}
        \eta_{n+1}(-\tau,t) &= \frac{1}{q_0}\bar x_2(\psi_2(\tau),t) - \vect g_2^\top(\tau)\vect\xi(t) \\
        &\hspace{1cm} - \int_0^{\psi_2(\tau)} g_2(\tau,z)\bar x_2(z,t)\,\dd z, && \tau\in[0,\tau_2]. \nonumber
    \end{align}
\end{subequations}
Specifically, \eqref{eq:flat_trafo_compact_ode} follows from inserting the solution of the Volterra integral equation \eqref{eq:flat_inteq2} into \eqref{eq:flat_HCCF_trafo_ODE}.
The map in \eqref{eq:flat_trafo_compact_pdep} (resp.\ \eqref{eq:flat_trafo_compact_pdem}) is obtained by using \eqref{eq:flat_x10_predicted} (resp.\ \eqref{eq:flat_x10_delayed}) as well as the solution $\vect\xi(t+\phi_1(z))$ (resp.\ $\vect\xi(t-\phi_2(z))$) of \eqref{eq:flat_inteq1} (resp.\ \eqref{eq:flat_inteq2}) in \eqref{eq:flat_HCCF_trafo_PDE}.
The partitioning\footnote{Note that $\vect g_1^\top(0)=\vect g_2^\top(0)+\frac{1}{q_0}\vect c^\top$, which allows to include $\tau=0$ in both \eqref{eq:flat_trafo_compact_pdep} and \eqref{eq:flat_trafo_compact_pdem}.} in \eqref{eq:flat_trafo_compact_pdep} and \eqref{eq:flat_trafo_compact_pdem} is done to better highlight that $\eta_{n+1}(\tau,t)$ is independent of $\bar x_1(z,t)$ for $\tau\in[-\tau_2,0]$ and independent of $\bar x_2(z,t)$ for $\tau\in[0,\tau_1]$, similar to the case of a hyperbolic PDE without an ODE depicted in Figure~\ref{fig:HCCFtrafo_n0}.
In \eqref{eq:flat_trafo_compact}, the explicit definition of $\mat G_0\in\Rset^{n\times n}$ as well as of the continuously differentiable functions therein, where $\vect g_0(z),\vect g_1(\tau),\vect g_2(\tau)\in\Rset^n$ and $g_1(\tau,z),g_2(\tau,z)\in\Rset$, is omitted due to their length and complexity.
The following lemma summarizes the results on the \gls{HCCF} state transformation.

\begin{lem}[\gls{HCCF} state transformation]
\label{lem:hccf_trafo}
    The map \eqref{eq:flat_trafo_compact} between the states $\vect\xi(t)$, $\bar{\vect x}(z,t)$, $z\in[0,1]$ of \eqref{eq:sys_transp} and $\vect\eta(t)$, $\eta_{n+1}(\tau,t)$, $\tau\in[-\tau_2,\tau_1]$ of the \gls{HCCF} \eqref{eq:flat_hccf} is bijective.
\end{lem}

The inverse map follows from the flatness-based parametrizations \eqref{eq:flat_parametrization_z0_xi} and \eqref{eq:flat_x_para} of $\vect\xi(t)$ and $\bar{\vect x}(z,t)$, respectively.
For that, repeatedly apply integration by parts as well as Cauchy's formula for repeated integration to get
\begin{align}
\label{eq:flat_uglyeq1}
    &\tint_{\ubar\tau}^{\bar\tau} f(\tau)y^{(i)}(t+\tau)\,\dd\tau \\
    &= \sum_{k=i}^{n-1} \int_{\ubar\tau}^{\bar\tau} (\bar\tau-\tau)^{k-i}f(\tau)\,\dd\tau \frac{(-1)^{k-i}}{(k-i)!} y^{(k)}(t+\bar\tau) \nonumber \\
    &+ \tint_{\ubar\tau}^{\bar\tau} \int_{\ubar\tau}^\tau (\tau-\sigma)^{n-i-1}f(\sigma)\,\dd\sigma \frac{(-1)^{n-i}}{(n-i-1)!} y^{(n)}(t+\tau)\,\dd\tau \nonumber
\end{align}
for $i=0,\dots,n-1$ and any continuous function $f(\tau)$, $-\tau_2\le\ubar\tau\le\tau\le\bar\tau\le\tau_1$.
In \eqref{eq:flat_x_para}, following a transformation of the integration variable, \eqref{eq:flat_uglyeq1} allows to replace integrals over $y^{(i)}(t+\tau)$, $i=0,\dots,n-1$, by ones over $y^{(n)}(t+\tau)=\eta_{n+1}(\tau,t)$.
Similarly, use
\begin{multline}
\label{eq:flat_uglyeq2}
    y^{(i)}(t+\tau) = \sum_{k=i}^{n-1} \frac{(\tau+\tau_2)^{k-i}}{(k-i)!} y^{(k)}(t-\tau_2) \\
    + \tint_{-\tau_2}^{\tau} \frac{(\tau-\sigma)^{n-i-1}}{(n-i-1)!} y^{(n)}(t+\sigma)\,\dd\sigma
\end{multline}
for $i=0,\dots,n-1$ and $\tau\in[-\tau_2,\tau_1]$ to express $y^{(i)}(t+\tau)$ in terms of the \gls{HCCF} state in \eqref{eq:flat_hccf_state}.
Thus, the right-hand sides of \eqref{eq:flat_parametrization_z0_x10} and \eqref{eq:flat_x_para} only depend on $\vect\eta(t)$ and $\eta_{n+1}(\tau,t)$.
This allows to write the flatness-based parametrizations as the transformation
\begin{subequations}
\label{eq:flat_trafo_compact_inv}
    \begin{align}
        \vect\xi(t) &= \mat H_0\vect\eta(t) + \int_{-\tau_2}^0 \vect h_0(\tau) \eta_{n+1}(\tau,t)\,\dd\tau \\
        \bar x_1(z,t) &= \eta_{n+1}(\phi_1(z),t) + \vect h_1^\top(z)\vect\eta(t) \\
        &\hspace{1cm} + \int_{-\tau_2}^{\phi_1(z)} h_1(z,\tau) \eta_{n+1}(\tau,t)\,\dd\tau, && z\in[0,1] \nonumber \\
        \bar x_2(z,t) &= q_0\eta_{n+1}(-\phi_2(z),t) + \vect h_2^\top(z)\vect\eta(t) \\
        &\hspace{1cm} + \int_{-\tau_2}^0 h_2(z,\tau) \eta_{n+1}(\tau,t)\,\dd\tau, && z\in[0,1], \nonumber
    \end{align}
\end{subequations}
which is the inverse of \eqref{eq:flat_trafo_compact}.
The lengthy definition of $\mat H_0\in\Rset^{n\times n}$ as well as of the continuously differentiable functions in \eqref{eq:flat_trafo_compact_inv}, where $\vect h_0(\tau),\vect h_1(z),\vect h_2(z)\in\Rset^n$ and $h_1(z,\tau),h_2(z,\tau)\in\Rset$, is omitted.


\renewcommand{\bibfont}{\normalfont}

\printbibliography

\begin{contributors}

    \contributor{Nicole Gehring}
        {Institute of Automatic Control and Control Systems Technology, Johannes Kepler University Linz, Austria}
        {nicole.gehring@jku.at}
        {}
        {Nicole Gehring received the Dipl.-Ing.\ degree in electrical engineering from Dresden University of Technology, Germany, in 2007 and the Dr.-Ing.\ degree in automatic control from Saarland University, Germany, in 2015. During her two-year stint in the industrial sector, she gained experience in the control of power plants and motor vehicles. As a Postdoc, she was with Technical University of Munich, Germany. She is currently with Johannes Kepler University Linz, Austria, where her research mainly focuses on the design of controllers and observers for linear distributed-parameter systems.}
        
    \contributor{Abdurrahman Irscheid}
        {Chair of Systems Theory and Control Engineering, Saarland University, Germany}
	{a.irscheid@lsr.uni-saarland.de}
	{}
        {Abdurrahman Irscheid received the B.Sc.\ degree in Mechatronics and the M.Sc.\ degree with honors in Computational Engineering of Technical Systems from Saarland University, Germany, in 2015 and 2017, respectively. He is currently working towards his Dr.-Ing.\ degree at the Chair of Systems Theory and Control Engineering at Saarland University.  His main research topics concern the controller and observer design for nonlinear distributed-parameter systems as well as control-theoretic methods for convergence in prescribed finite time.}

    \contributor{Joachim Deutscher}
        {Institute of Measurement, Control and Microtechnology, Ulm University, Germany}
        {joachim.deutscher@uni-ulm.de}
        {}
        {Joachim Deutscher received 
        the Dr.-Ing.\ and the Dr.-Ing.\ habil.\ degrees both in automatic control from Friedrich-Alexander-Universität Erlangen-Nürnberg (FAU), Germany, in 2003 and 2010, respectively. 
        In 2011 he was appointed Associate Professor at FAU and in 2017 he became a Professor at the same university. Since April 2020 he is a Full Professor at the Institute of Measurement, Control and Microtechnology at Ulm University. His research interests include control of distributed-parameter and multi-agent systems as well as data-based control with applications in mechatronics. 
        At present he serves as Associate Editor for Automatica.}

    \contributor{Frank Woittennek}
        {Institute of Automation and Control Engineering, UMIT TIROL -- Private University for Health Sciences and Health Technology, Austria}
        {frank.woittennek@umit-tirol.at}
        {}
        {Frank Woittennek received the Dr.-Ing.\ degree in electrical engineering from Technische Universität Dresden (TU Dresden), Germany, in 2007. Since 2015 he has been a Full Professor and the Head of the Institute of Automation and Control Engineering at UMIT TIROL. His current research interests include analysis, identification, control, and observer design for distributed parameter systems and nonlinear systems as well as applications in the fields of mechatronics, robotics, process engineering, and energy systems.}

    \contributor{Joachim Rudolph}
        {Chair of Systems Theory and Control Engineering, Saarland University, Germany}
        {j.rudolph@lsr.uni-saarland.de}
        {}
        {Joachim Rudolph received the doctorate degree from Université Paris XI, Orsay, France, in 1991, and the Dr.-Ing.\ habil.\ degree from Technische Universität Dresden, Germany, in 2003. Since 2009, he has been the Head of the Chair of Systems Theory and Control Engineering at Saarland University, Saarbrücken, Germany. His current research interests include controller and observer design for nonlinear and infinite dimensional systems, algebraic systems theory, and the solution of demanding practical control problems.}
        
\end{contributors}

\end{document}